\documentclass[11pt]{article}
\usepackage[margin=1in]{geometry}
\usepackage[T1]{fontenc}
\usepackage{lmodern,microtype,amsmath,amssymb,amsthm,mathtools}
\usepackage{graphicx,booktabs,fvextra,float,placeins,needspace,algorithm,algpseudocode}
\newfloat{listing}{htbp}{lol}
\floatname{listing}{Listing}
\usepackage{array,longtable}
\usepackage[round,authoryear]{natbib}
\usepackage[commandRef=cref]{proof-at-the-end}
\usepackage{xr-hyper}
\usepackage[hidelinks]{hyperref}
\usepackage[capitalize,nameinlink,noabbrev]{cleveref}
\ifdefined\CombinedStandaloneSupplement
\fi
\theoremstyle{definition}

\newtheorem{definition}{Definition}
\theoremstyle{plain}
\crefname{theorem}{Theorem}{Theorems}
\crefname{proposition}{Proposition}{Propositions}
\crefname{lemma}{Lemma}{Lemmas}
\crefname{corollary}{Corollary}{Corollaries}
\crefname{assumption}{Assumption}{Assumptions}
\crefname{definition}{Definition}{Definitions}
\crefname{algorithm}{Algorithm}{Algorithms}
\crefname{listing}{Listing}{Listings}
\DeclareMathOperator{\E}{E}
\DeclareMathOperator{\Var}{Var}
\DeclareMathOperator{\Cov}{Cov}

\DeclareMathOperator{\TV}{TV}
\newcommand{\R}{\mathbb R}

\ifdefined\CombinedInlineProofs
\pgfkeys{/prAtEnd/combined main/.style={normal}}
\else
\pgfkeys{/prAtEnd/combined main/.style={end,restate,no link to proof}}
\fi
\newcommand{\CombinedResultCheck}[2]{%
\typeout{COMBINED-RESULT label=#1 kind=#2 number=\csname the#2\endcsname\space zone=\CombinedProofZone}}
\newcommand{\CombinedProofCheck}[1]{%
\typeout{COMBINED-PROOF label=#1 zone=\CombinedProofZone}}

\newcommand{\AppEndpointPercent}{4.6\%}
\newcommand{\AppPpcMaximum}{0.23}
\newcommand{\OrdPpcMaximum}{0.48}
\newcommand{\AppPpcSequence}{0.31}
\newcommand{\OrdPpcSequence}{0.52}

\newcommand{\AppGainMean}{3.35}
\newcommand{\AppGainLower}{2.73}
\newcommand{\AppGainUpper}{3.99}
\newcommand{\OrdGainMean}{3.00}
\newcommand{\OrdGainLower}{2.00}
\newcommand{\OrdGainUpper}{4.01}

\newcommand{\CombinedArticleTitle}{Positive, monotone, and convex Gaussian process regression with Laplacian eigenfunctions}
\title{\CombinedArticleTitle}
\author{Andreas Kryger Jensen\\Section of Biostatistics, Department of Public Health\\University of Copenhagen}
\date{}
\hypersetup{pdftitle={Positive, monotone, and convex Gaussian process regression with Laplacian eigenfunctions},pdfauthor={Andreas Kryger Jensen}}

\begin{document}
\maketitle
\def\CombinedProofZone{main}
\begin{abstract}
We study how to condition an existing Hilbert space Gaussian process
(HSGP) expansion on positivity, monotonicity, or convexity throughout
an interval. The expansion has Gaussian coefficients whose variances
are given by the kernel's spectral density. At fixed covariance and
noise parameters, we retain this coefficient law and represent each
complete interval event as a threshold on one coefficient. Integrating
out that coefficient gives a strongly log-concave marginal posterior
with a normal survival weight, while conditional truncated normal
sampling recovers the full constrained law. The construction also
covers joint restrictions within the finite expansion. For individual
restrictions, we characterize the thresholds through polynomial
equations and bound posterior error from conservative numerical
evaluation in total variation. Under spectral regularity, the
conditioned finite posteriors converge in quadratic Wasserstein
distance to the corresponding conditioned underlying GP posteriors as
the computational interval expands and higher frequencies are
retained. This convergence controls uniform discrepancies in the
response and derivatives through the constrained order. Mercury lidar
and leaf light-response applications illustrate the consequences for
derivative inference.
\end{abstract}
\section{Introduction}\label{sec:introduction}

Scientific information about a response often concerns its sign or how
it changes over an interval. Positivity restricts the response itself,
while monotonicity and convexity restrict its slope and curvature when
the corresponding derivatives exist. These properties need not be
visible in noisy observations, so a Gaussian process (GP) that describes
the data closely can still assign probability to curves that conflict
with the proposed scientific model. We therefore seek a posterior law
that imposes the restriction throughout a specified interval while
retaining joint uncertainty about the response and its derivatives.

The Hilbert space GP (HSGP) approximation of \citet{solin2020hsgp}
represents the process by a finite expansion in Laplacian eigenfunctions
on an interval extending beyond the region of interest. This extension,
called padding, is used to reduce boundary effects, while the rank,
or number of retained functions, controls the finest variation represented
for a given computational interval \citep{riutort2023practicalhsgp}.
The basis functions are independent of the covariance kernel and its
parameters. The kernel enters through spectral values that determine
Gaussian coefficient variances, allowing the same basis and its analytic
derivatives to be reused across covariance specifications.
Shape restrictions, however, concern
the entire curve obtained by summing the oscillatory basis functions.
Their validity is not determined by the signs of the individual
coefficients, and checking a finite set of locations can miss a
violation between them. The question addressed here is how to
condition this spectral model on the complete interval restriction
while retaining its basis and kernel-derived Gaussian coefficient law.

Our construction begins with positivity. A random intercept shifts
the whole curve above zero when it exceeds the negative minimum of
the spectral part. The same argument extends to monotonicity and
convexity whenever the retained span contains a direction with a
uniformly positive selected derivative, because a sufficiently large
coefficient along that direction makes the full curve satisfy the
restriction for any fixed values of the remaining coefficients. Thus the
continuum event becomes a lower bound on one Gaussian coefficient,
with a threshold determined by the remaining coefficients. Under
Gaussian observations, integrating out that coefficient gives an
explicit marginal posterior weighted by a normal survival probability.
A subsequent draw from its truncated conditional normal restores
the full conditioned law. This representation makes continuum
conditioning available within the HSGP expansion without
changing the probability model through a nonlinear transformation
of the curve.

\subsection{Related work and contributions}

Conditioning Gaussian processes on inequalities is an established
approach to shape-constrained regression. The survey of
\citet{swiler2020survey} distinguishes constraints imposed at selected
locations from constructions that restrict an entire finite-dimensional
function space. In particular, \citet{daveiga2012inequality} formulate
bound, monotonicity, and convexity restrictions as conditioning events
for a GP. To compute constrained predictive means, they replace
restrictions over an interval by inequalities at finitely many selected
locations and evaluate or approximate moments of the resulting
truncated multivariate normal law. Their later work develops a strategy
for selecting these locations \citep{daveiga2020linear}. The target of
our construction is also a conditioned Gaussian law. The distinction
is that, within a retained HSGP expansion, we characterize the
restriction over every location in the interval before computing its
posterior law.

Global derivative restrictions and derivative inference have precedents
in Bayesian regression, where piecewise-linear isotonic models permit
flat and increasing segments \citep{neelon2004isotonic}, while smooth
parameterizations and regression splines give other monotone priors
\citep{shively2009monotone}. Ordered Bernstein coefficients combine
global monotonicity with variable selection or clustering
\citep{curtis2011bernstein,wilson2020monotone}, with the aerosol analysis
of \citet{wilson2020monotone} illustrating derivative inference through
the derivative of a monotone differential-pressure curve.
For convex regression, maxima of random affine functions address
multivariate inputs and permit nonsmooth surfaces
\citep{hannah2011convex}, for which support and posterior consistency
results are available. Integrating squared GPs provides smooth
shape restrictions \citep{lenk2017shape}, with spectral constructions
also available for grouped data \citep{lenk2024hierarchical}.
These approaches establish that global shape information and
derivative inference can be incorporated in Bayesian regression in
several ways. They also make the choice of probability model
substantive: transforming or integrating a GP generally gives a
different law from conditioning a specified Gaussian model on a
shape event.

The closest whole-interval constructions condition finite Gaussian
expansions using bases chosen to express global shape restrictions as coefficient
inequalities \citep{maatouk2017constraints}. These constraints hold
throughout the domain, and subsequent work develops general linear
inequalities, noisy observations, multiple constraints, and smooth
bases \citep{lopera2018finite,lopera2020noisy,maatouk2025bayesian}.
Within this line of work, \citet{ray2020shaperestricted} combine
smooth relaxations of coefficient constraints with elliptical slice
sampling and circulant embedding, while Gaussian simulation on
correlated subdomains addresses larger problems
\citep{maatouk2025largescale}. Our construction shares the aim of
conditioning a finite Gaussian model on a global restriction. Its
distinctive task is to characterize that restriction in the
Laplacian--Dirichlet spectral space while retaining its kernel-derived
Gaussian coefficient law. The oscillatory basis requires a different
description of the constraint. Laplacian spectral expansions themselves
also have precedents, including their use for boundary conditions
discussed by \citet{swiler2020survey}.

Other GP approaches impose derivative signs at virtual locations
\citep{riihimaki2010monotonicity,agrell2019operators}, use randomized
optimization for constrained inference \citep{zhang2026rlrto}, or
project GP curves onto the monotone-function space
\citep{lin2014projection}. Virtual-location restrictions provide a
finite approximation to an interval constraint, whereas projection
defines the fitted curve by a separate map. \cref{tab:literature-position}
locates the present construction among the approaches most directly
related to its conditioning target. The table separates whether a
restriction holds throughout the finite model's interval from whether
that finite model approximates a conditioned underlying GP.

\begin{table}[htbp]
\centering
\small
\setlength{\tabcolsep}{4pt}
\begin{tabular}{@{}
 >{\raggedright\arraybackslash}p{0.18\textwidth}
 >{\raggedright\arraybackslash}p{0.27\textwidth}
 >{\raggedright\arraybackslash}p{0.25\textwidth}
 >{\raggedright\arraybackslash}p{0.24\textwidth}@{}}
\toprule
Approach & Model and imposed restriction & Computed object & Interval scope\\
\midrule
Da Veiga--Marrel
 & GP conditioned on inequalities at selected locations
 & Constrained predictive moments from a truncated Gaussian law
 & The selected-location event approximates the interval event\\
Maatouk--Bay and extensions
 & Finite Gaussian expansion in a basis chosen for coefficient inequalities
 & Conditioned coefficient law and predictive summaries
 & Coefficient inequalities ensure shape throughout the finite expansion's domain\\
Present paper
 & HSGP expansion with its kernel-derived Gaussian coefficient law conditioned on an interval event
 & Survival-weighted marginal law and conditional reconstruction of the full coefficient law
 & Exact interval event at finite rank and a conditioned-GP limit for individual restrictions\\
\bottomrule
\end{tabular}
\caption{Conditioning targets and interval guarantees for related approaches.
The descriptions concern the finite models used for computation. The
conditioned-GP limit in the last row requires the spectral assumptions
and increasing padding and frequency cutoff stated in
\cref{thm:approximation}.}
\label{tab:literature-position}
\end{table}

The first contribution is a common representation for positivity,
monotonicity, and convexity, together with explicit directions and
sufficient rank and padding conditions for each case. Sign reversal
gives the corresponding nonpositive, nonincreasing, and concave
models. The representation also applies to joint restrictions whenever
the retained span contains a direction with the required signs at all
selected derivative orders. We give conditions on the HSGP direction
coefficients and threshold reductions covering all 26 combinations
without opposing constraints.

To evaluate the thresholds for the individual constraints, a cosine
transformation reduces the search for stationary
locations to polynomial equations, with endpoints, repeated roots,
and constant cases included. These characterizations make the
thresholds amenable to global evaluation. The resulting marginal
posterior is strongly log-concave in whitened coordinates, and a
total variation bound quantifies the change in its reconstructed
law when a conservative numerical threshold is used. The
computational cost is a global maximization at each target evaluation,
so the representation offers an alternative to bases with direct
coefficient inequalities, rather than a general efficiency advantage.

The second contribution connects finite continuum conditioning to
conditioning the underlying GP. Covariance approximation alone does
not settle this question because the indicator of a sign event is
discontinuous at curves that touch its boundary. For each individual
constraint, we establish a null boundary and control function-space
moments, obtaining quadratic Wasserstein convergence of the
conditioned posterior as padding and frequency cutoff grow. Under
the stated spectral regularity, convergence holds in the uniform
norm for the response and all derivatives up to the constrained
order. It therefore also gives uniform convergence of their posterior
means and covariance functions. The result includes Mat\'ern
smoothness greater than zero for positivity, greater than one for
monotonicity, and greater than two for convexity.

Two applications show why retaining joint uncertainty about the
response and its derivatives matters. Under the differential
absorption lidar approximation, mercury concentration is proportional
to the range derivative of a latent negative log signal ratio, so
nonnegative concentration requires that curve to be nondecreasing.
For leaf light-response data, concavity instead encodes a nonincreasing
marginal gain in assimilation as irradiance increases, without
requiring assimilation itself to increase throughout the range.
These analyses illustrate how a scientifically specified shape
restriction changes derivative inference.

This paper is organized as follows. \cref{sec:methodology} introduces the
spectral model and its conditioning events, followed by the Gaussian
coefficient construction for positivity, monotonicity, convexity, and
their joint combinations in \cref{sec:methods}. \cref{sec:theory} studies
posterior geometry, numerical conditioning error, and convergence to the
conditioned underlying GP, and \cref{sec:computation} describes posterior
computation. \cref{sec:applications} presents the mercury lidar and leaf
light-response applications, followed by the discussion in
\cref{sec:discussion}.
\ifdefined\CombinedInlineProofs\else
All proofs are given in the supplementary material.
\fi

Stan implementations of the monotone and convex models, together with
R code for reproducing the two applications, are available from the
author's GitHub repository.

\section{Methods}\label{sec:methodology}
\label{sec:model}

The Hilbert space Gaussian process (HSGP) construction of
\citet{solin2020hsgp} approximates a stationary GP by a finite
expansion in Laplacian eigenfunctions with Gaussian coefficients.
The domain determines the basis functions, while the covariance
kernel determines the coefficient variances. Changing the kernel or
its parameters therefore leaves the basis and its derivatives
unchanged, allowing their values at observation and prediction
locations to be computed once and reused
\citep{riutort2023practicalhsgp}. For shape-constrained regression,
a further advantage is that the response, slope, and curvature are
linear functions of the same coefficients. We can therefore express
all three restrictions in a common coefficient space and obtain
their conditioned laws from the ordinary Gaussian regression posterior.

Let $k(u)$ be a real stationary covariance function with a spectral
density, where $u=x-x'$ is the separation between inputs. We write
$S_k:\R\to[0,\infty)$ for its nonnegative, even, integrable spectral
density and use the Fourier convention
\[
 k(u)=\frac1{2\pi}\int_\R e^{\mathrm i\omega u}S_k(\omega)\,d\omega,
 \qquad u\in\R.
\]
Here $\omega$ is angular frequency, and $S_k(\omega)$ is the spectral
density associated with the covariance function $k$ through this
Fourier relation, evaluated at frequency $\omega$. The HSGP expansion
uses this density to assign variances at a discrete set of frequencies
determined by the Laplacian on a bounded interval.

Fix a center $c\in\R$, a half-width $L>0$, an integer $m\geq1$,
and a compact interval $I=[x_-,x_+]$ of positive length inside
$(c-L,c+L)$. On the padded interval $[c-L,c+L]$, the negative
Laplacian with zero boundary values has eigenvalues $\omega_j^2$
and orthonormal eigenfunctions $\phi_{j,L}$. Using the spectral
density $S_k$, these functions define the
finite Gaussian expansion
\begin{equation}\label{eq:finite-model}
 \begin{aligned}
 \phi_{j,L}(x)&=L^{-1/2}\sin(\omega_j(x-c+L)),
 &\omega_j&=j\pi/(2L),\\
 f_{\boldsymbol a}(x)&=a_0+\sum_{j=1}^ma_j\phi_{j,L}(x),
 &a_0&\sim N(\mu_0,\sigma_0^2),\quad a_j\sim N(0,S_k(\omega_j)).
 \end{aligned}
\end{equation}
The coefficients are independent, with $\sigma_0>0$ and
$S_k(\omega_j)>0$ at each retained frequency, so the finite coefficient
law is nondegenerate.

Under this prior, the covariance is
\[
 \Cov(f_{\boldsymbol a}(x),f_{\boldsymbol a}(x'))
 =\sigma_0^2+\sum_{j=1}^m S_k(\omega_j)
                         \phi_{j,L}(x)\phi_{j,L}(x').
\]
The sum approximates the stationary covariance $k(x-x')$, while
$\sigma_0^2$ is the contribution of the random intercept. Because
the eigenfunctions vanish at the padded endpoints, the finite
covariance is generally nonstationary. Defining the basis on an
interval larger than $I$ places this artificial boundary outside
the region of inference without requiring the response to vanish
at the endpoints of $I$. The intercept permits uncertainty in the
overall response level without changing slope or curvature.

\label{sec:finite-model}
For observations $\boldsymbol y\in\R^n$ at $x_i\in I$,
the Gaussian observation model is
\[
 \boldsymbol Y\mid\boldsymbol a\sim
 N\bigl((f_{\boldsymbol a}(x_1),\ldots,f_{\boldsymbol a}(x_n))^T,
        \boldsymbol\Sigma\bigr),
\]
where the noise covariance $\boldsymbol\Sigma$ is positive
definite and the errors are independent of the coefficients.
For given covariance and noise parameters and a chosen basis, we study
the coefficient posterior conditional on the observed responses and
their locations.

Write $\boldsymbol a=(a_0,\ldots,a_m)^T$. If $\boldsymbol X$ has rows
$(1,\phi_{1,L}(x_i),\ldots,\phi_{m,L}(x_i))$, set
$\boldsymbol m_0=(\mu_0,0,\ldots,0)^T$ and
$\boldsymbol V_0=\operatorname{diag}(\sigma_0^2,S_k(\omega_1),\ldots,S_k(\omega_m))$.
The coefficients and observations then have the joint Gaussian distribution
\[
 \begin{pmatrix}\boldsymbol a\\\boldsymbol Y\end{pmatrix}
 \sim N\left(
 \begin{pmatrix}\boldsymbol m_0\\\boldsymbol X\boldsymbol m_0\end{pmatrix},
 \begin{pmatrix}
  \boldsymbol V_0 & \boldsymbol V_0\boldsymbol X^T\\
  \boldsymbol X\boldsymbol V_0 &
  \boldsymbol X\boldsymbol V_0\boldsymbol X^T+\boldsymbol\Sigma
 \end{pmatrix}
 \right).
\]
Here $\boldsymbol\Sigma$ is the noise covariance, while
$\boldsymbol X\boldsymbol V_0\boldsymbol X^T+\boldsymbol\Sigma$
is the marginal covariance of the observations. Conditioning this joint
distribution on $\boldsymbol Y=\boldsymbol y$ gives the ordinary Gaussian posterior
\begin{equation}\label{eq:gaussian-posterior}
 \begin{aligned}
 \boldsymbol a\mid\boldsymbol y&\sim N(\boldsymbol m,\boldsymbol V),\\
 \boldsymbol m&=\boldsymbol m_0+\boldsymbol V_0\boldsymbol X^T
       (\boldsymbol X\boldsymbol V_0\boldsymbol X^T+\boldsymbol\Sigma)^{-1}
       (\boldsymbol y-\boldsymbol X\boldsymbol m_0),\\
 \boldsymbol V&=\boldsymbol V_0-\boldsymbol V_0\boldsymbol X^T
       (\boldsymbol X\boldsymbol V_0\boldsymbol X^T+\boldsymbol\Sigma)^{-1}
       \boldsymbol X\boldsymbol V_0.
 \end{aligned}
\end{equation}
These equations specify the Gaussian law that we
will condition on shape, with factorizations and linear solves used
in its numerical preparation.

We treat positivity, monotonicity, and convexity as sign restrictions
of orders zero, one, and two. Write $C^0(I)=C(I)$ and
$\frac{\mathrm{d}^0f}{\mathrm{d}x^0}=f$. For $r\in\{0,1,2\}$ we condition the ordinary
posterior on the function-space event
\begin{equation}\label{eq:shape-event}
 \mathcal C_r=\left\{f\in C^r(I):
                 \min_{x\in I}\frac{\mathrm{d}^rf}{\mathrm{d}x^r}(x)\geq0\right\}.
\end{equation}
The zero order gives a nonnegative response, the first gives
nondecreasing regression, and the second gives convex regression, with
$\mathcal C_r$ also denoting its explicitly identified preimage in
coefficient space. The restriction requires nonnegativity but does
not impose a fixed positive lower bound on the response, slope, or
curvature.

Shape information can equivalently be incorporated into the prior.
Provided that $\mathcal C_r$ has positive prior probability, we may
restrict the Gaussian prior to curves in $\mathcal C_r$ and then update
it using the observation likelihood. This gives the same posterior as
first performing ordinary Gaussian updating and then conditioning on
$\mathcal C_r$, because the prior normalizing constant cancels in
Bayes' formula.

\subsection{Continuum constraints through a Gaussian coefficient}\label{sec:methods}

To express a continuum restriction through one coefficient, we first
choose a function in the retained span whose selected derivative,
including the function itself at order zero, is positive throughout $I$.
A sufficiently large coefficient along this function then makes the
selected quantity nonnegative everywhere on the interval, whatever
the values of the remaining coefficients. Since an invertible change
of coordinates preserves every admissible coefficient vector in the
finite space, the restriction can be represented by a single lower
bound without changing the underlying Gaussian law. The following
result integrates that coefficient and reconstructs it conditionally,
using the standard normal distribution function $\Phi$ and survival
function $\Phi^{\mathrm c}(x)=1-\Phi(x)$.

\begin{theoremEnd}[combined main,category=finite]{theorem}[Gaussian conditioning under a continuum sign restriction]
\CombinedResultCheck{thm:finite}{theorem}\label{thm:finite}
Under the finite Gaussian model in \cref{sec:finite-model}, let
$\boldsymbol a=(a_0,\ldots,a_m)^T$ denote the coefficient vector and fix
$r\in\{0,1,2\}$. Suppose an invertible linear change of coordinates
$\boldsymbol a\mapsto(\boldsymbol g,\gamma)\in\R^m\times\R$
gives $f_{\boldsymbol a}=f_{\boldsymbol g}+\gamma h_r$.
Here $h_r$ is a fixed function in the retained span, chosen to have a
uniformly positive $r$th derivative on $I$, and $\gamma$ is its scalar
coefficient. The curve $f_{\boldsymbol g}$ depends linearly on the
remaining coordinates $\boldsymbol g$. Assume these functions belong
to $C^r(I)$ with
$\min_{x\in I}\frac{\mathrm{d}^rh_r}{\mathrm{d}x^r}(x)>0$.
In these coordinates, the ordinary Gaussian posterior has the form
\begin{equation}\label{eq:conditional}
 \boldsymbol g\mid\boldsymbol y\sim N(\boldsymbol m_g,\boldsymbol V_g),
 \qquad \gamma\mid\boldsymbol g,\boldsymbol y\sim N(\mu(\boldsymbol g),s^2).
\end{equation}
Here $\boldsymbol V_g$ is positive definite, $\mu$ is affine, and
$s>0$ is constant in $\boldsymbol g$. The finite, attained threshold
\begin{equation}\label{eq:threshold}
 \tau_r(\boldsymbol g)=\max_{x\in I}
 \frac{-\frac{\mathrm{d}^rf_{\boldsymbol g}}{\mathrm{d}x^r}(x)}
      {\frac{\mathrm{d}^rh_r}{\mathrm{d}x^r}(x)}
\end{equation}
satisfies $f_{\boldsymbol a}\in\mathcal C_r$ exactly when
$\gamma\geq\tau_r(\boldsymbol g)$. This event has positive ordinary
posterior probability and a null boundary in coefficient space.
Its marginal density for $\boldsymbol g$ with
respect to Lebesgue measure on $\R^m$ is
\begin{equation}\label{eq:marginal}
 p(\boldsymbol g\mid\boldsymbol y,\mathcal C_r)=
 \frac{p(\boldsymbol g\mid\boldsymbol y)
 \Phi^{\mathrm c}((\tau_r(\boldsymbol g)-\mu(\boldsymbol g))/s)}
 {P(f_{\boldsymbol a}\in\mathcal C_r\mid\boldsymbol y)}.
\end{equation}
Given this draw of $\boldsymbol g$, draw $\gamma$ from the conditional normal in
\cref{eq:conditional} truncated below at $\tau_r(\boldsymbol g)$ and
apply the inverse coefficient map. Equivalently, for an independent
$U\sim\operatorname{Uniform}(0,1)$,
\begin{equation}\label{eq:reconstruction}
 \gamma=\mu(\boldsymbol g)-s\Phi^{-1}
 \bigl(U\Phi^{\mathrm c}((\tau_r(\boldsymbol g)-\mu(\boldsymbol g))/s)\bigr).
\end{equation}
This reconstructs the full conditioned coefficient law, under which
$\min_I\frac{\mathrm{d}^rf_{\boldsymbol a}}{\mathrm{d}x^r}>0$ almost surely.
\end{theoremEnd}

\begin{proofEnd}
\CombinedProofCheck{thm:finite}
Fix $r$, $I$, the invertible map, and the ordinary posterior under
the finite Gaussian model in \cref{sec:finite-model} and the
positive-direction hypotheses of
\cref{thm:finite}, and write
$b_r=\min_{x\in I}\frac{\mathrm{d}^rh_r}{\mathrm{d}x^r}(x)>0$.
The role of the positive direction is to turn every
derivative inequality into a lower bound on the same scalar coefficient:
division by $\frac{\mathrm{d}^rh_r}{\mathrm{d}x^r}$ gives the threshold equivalence,
and continuity on compact $I$, together with $b_r>0$, makes
\cref{eq:threshold} finite and attained. Since the invertible map
preserves positive definiteness of the posterior covariance, Gaussian conditioning
\citep[Section 2.2]{rasmussen2006gpml} gives
\cref{eq:conditional}, including a positive Schur complement $s^2$.

Integrating the restricted joint Gaussian density over the scalar
coefficient now leaves its conditional survival probability as a
weight on $\boldsymbol g$, which proves \cref{eq:marginal}.
This weight is strictly positive for every $\boldsymbol g$, so the
conditioning event has positive probability. Given a draw from the
weighted marginal, conditioning the Gaussian law of
$\gamma\mid\boldsymbol g,\boldsymbol y$ on
$\gamma\geq\tau_r(\boldsymbol g)$ gives the stated truncated normal,
whose survival-function inverse is \cref{eq:reconstruction}.
Applying the inverse coefficient map therefore recovers the original
coefficient law conditional on the sign restriction.

The minimum derivative is continuous in the coefficients, so coefficient
vectors with a positive minimum lie in the event's interior and those
with a negative minimum lie in its exterior. Conditional on
$\boldsymbol g$, the nondegenerate normal assigns zero
mass to equality at the threshold. At equality an attained maximizing
location has zero derivative, whereas above it
\[
 \min_I\frac{\mathrm{d}^rf_{\boldsymbol a}}{\mathrm{d}x^r}
 \geq(\gamma-\tau_r(\boldsymbol g))b_r>0.
\]
At equality, an arbitrarily small increase or decrease in $\gamma$
moves to either side of the event, identifying its coefficient boundary
as precisely the threshold graph.
Its zero Gaussian probability implies that the reconstructed law lies
strictly above the graph almost surely, proving the strict-derivative
assertion.
\end{proofEnd}

The Gaussian tail in \cref{eq:marginal} is the conditional probability
that the separated coefficient is large enough to satisfy the
constraint, so values of $\boldsymbol g$ receive less posterior weight
when they require an unlikely value of that coefficient. Conditional
reconstruction retains the resulting dependence on $\boldsymbol g$ and
therefore gives joint draws of the curve and its derivatives. This
dependence is inherited from the original Gaussian model and must be
transported through the coefficient map, since assigning a new
independent prior to the transformed coordinates would change the law.

A related conditioning structure appears in GP regression with censored
observations, where a Gaussian posterior density is weighted by the
conditional probability of the censoring event
\citep{gormhoffmann2025censored}. In that setting, inequalities describe
censored measurements at finitely many locations, whereas the threshold
here represents a restriction on the latent curve throughout an interval.

The null-boundary statement in \cref{thm:finite} means that, under
its assumptions, the conditioned law assigns probability zero to curves
whose minimum response or selected derivative equals zero. Thus a draw
from this law has a strictly positive minimum almost surely, although
these minima can be arbitrarily close to zero. For monotonicity,
positivity of the derivative at every point is stronger than strict
ordering of function values, which alone does not require a positive
derivative everywhere. Nearly flat regions therefore remain possible,
but the analyticity of the finite expansion means that assigning
positive probability to exact plateaus or linear segments of positive
length requires a different prior.

The random intercept supplies a constant direction for positivity,
whereas monotonicity and convexity require directions with positive
slope and curvature. For each choice, evaluating the threshold still
requires finding a global extremum over $I$. On the padded interval of
length $2L$, the zero boundary conditions require
$\sin(2L\omega_j)=0$, giving
$\omega_j=j\pi/(2L)=j\omega_1$ for $j\geq1$. The basis and its
derivatives therefore involve sine and cosine harmonics of a common
fundamental frequency.
We introduce
\[
 \theta(x)=\omega_1(x-c+L),\qquad w(x)=\cos\theta(x),
\]
because each cosine harmonic becomes a polynomial in $w$, while each
sine harmonic becomes $\sin\theta$ times a polynomial in $w$, through
identities for the Chebyshev polynomials in \cref{def:chebyshev}.
These identities allow the stationary locations needed for the
thresholds to be characterized by polynomial equations.

Since $I$ lies strictly inside the padded domain, $0<\theta(x)<\pi$
and $\mathrm{d}w/\mathrm{d}x=-\omega_1\sin\theta(x)<0$ throughout $I$. Thus $w$ maps
$I$ one-to-one onto a compact interval in $(-1,1)$, reversing the order
of its endpoints. Its nonzero derivative ensures that interior
stationary locations correspond in the two coordinates, so the change
of variable preserves the complete search for a continuum extremum.

\begin{definition}[Chebyshev polynomials]\label{def:chebyshev}
The Chebyshev polynomials of the first kind, $T_j$, and second kind,
$U_j$, are defined for $w\in\R$ by the initial values and recurrences
\[
 \begin{aligned}
 T_0(w)&=1,\quad T_1(w)=w,
 & T_{j+1}(w)&=2wT_j(w)-T_{j-1}(w),\\
 U_0(w)&=1,\quad U_1(w)=2w,
 & U_{j+1}(w)&=2wU_j(w)-U_{j-1}(w),
 \end{aligned}
\]
where the recurrences hold for integers $j\geq1$.
\end{definition}

For $w=\cos\theta$, these polynomials satisfy the standard trigonometric
identities \citep[Equations (18.5.1) and (18.5.2)]{nist2026dlmf}
\[
 T_j(w)=\cos(j\theta),\qquad
 \sin\theta\,U_{j-1}(w)=\sin(j\theta),\qquad j\geq1.
\]
In particular, the basis and its first two derivatives satisfy
\[
 \phi_{j,L}(x)=\frac{\sin\theta(x)}{\sqrt L}\,U_{j-1}(w(x)),\quad
 \frac{\mathrm{d}\phi_{j,L}}{\mathrm{d}x}(x)=\frac{j\omega_1}{\sqrt L}\,T_j(w(x)),\quad
 \frac{\mathrm{d}^2\phi_{j,L}}{\mathrm{d}x^2}(x)=-\omega_j^2\phi_{j,L}(x).
\]
The second derivative therefore shares the sine factor in the first
identity. For positivity, the polynomial
representing the first derivative identifies the stationary locations
of the curve. With the directions chosen below, monotonicity gives a
ratio of polynomials with a positive denominator, while convexity gives
a polynomial after the common sine factor cancels. Differentiating
these expressions reduces their stationary-point conditions to
polynomial equations. Evaluating the interval
endpoints and these locations determines the global thresholds, with
constant cases handled explicitly in the corollaries.

The following three corollaries apply \cref{thm:finite} to positivity,
monotonicity, and convexity, specifying a suitable direction and the
corresponding continuum threshold for each restriction.

\begin{theoremEnd}[combined main,category=positive]{corollary}[Positivity in the Laplacian expansion]
\CombinedResultCheck{cor:positive}{corollary}\label{cor:positive}
Under the finite Gaussian model in \cref{sec:finite-model}, suppose
$m\geq1$ and $I\subset(c-L,c+L)$. Set
$h_0=1$, $\boldsymbol g=(a_1,\ldots,a_m)^T$, and
$f_{\boldsymbol g}=\sum_{j=1}^ma_j\phi_{j,L}$.
Then $f_{\boldsymbol a}=f_{\boldsymbol g}+a_0h_0$, with inverse
obtained by reinserting $a_0$.

For $\theta(x)=\omega_1(x-c+L)$, $w(x)=\cos\theta(x)$, and
$J=[w(x_+),w(x_-)]\subset(-1,1)$, define
$P_{0,\boldsymbol g}(w)=\sum_{j=1}^mja_jT_j(w)$. The threshold is
\begin{equation}\label{eq:positive-threshold}
 \tau_0(\boldsymbol g)=-\min_{x\in I}f_{\boldsymbol g}(x).
\end{equation}
\cref{thm:finite} applies with scalar coefficient $\gamma=a_0$.

The threshold is attained at a physical endpoint or a location whose
cosine coordinate is an interior real root in $J$ of
$P_{0,\boldsymbol g}$. This stationary polynomial has degree at most
$m$ when nonzero. All root multiplicities and reduced degrees are
included. If the stationary polynomial vanishes identically, the
curve $f_{\boldsymbol g}$ is constant and either endpoint suffices.
\end{theoremEnd}

\begin{proofEnd}
\CombinedProofCheck{cor:positive}
Use the finite Gaussian model in \cref{sec:finite-model} and its compact
physical interval. Separating the intercept is an invertible permutation
of the coefficients, with $\min_Ih_0=1$, so \cref{thm:finite} at
order zero gives \cref{eq:positive-threshold} and the conditioned law.
To locate the minimum over the whole interval, use the Chebyshev
identities to write
\[
 f_{\boldsymbol g}(x)=\frac{\sqrt{1-w(x)^2}}{\sqrt L}
             \sum_{j=1}^ma_jU_{j-1}(w(x)),
 \qquad
 \frac{\mathrm{d}f_{\boldsymbol g}}{\mathrm{d}x}(x)=
             \frac{\omega_1}{\sqrt L}P_{0,\boldsymbol g}(w(x)).
\]
Because the cosine is strictly decreasing for $0<\theta(x)<\pi$,
interior stationary locations correspond exactly to interior roots of
the displayed polynomial. The extreme value theorem and Fermat's
theorem therefore give the endpoint and stationary candidates, with
finitely many distinct roots, including repeated roots, when the
polynomial is nonzero and has degree at most $m$. If it vanishes
identically, the derivative is zero throughout $I$, so the mean value
theorem gives constancy. Although the intercept is
independent of the spectral coefficients in the prior, reconstruction
uses its conditional posterior given $\boldsymbol g$ and the data.
\end{proofEnd}

\begin{theoremEnd}[combined main,category=monotone]{corollary}[Monotonicity in the Laplacian expansion]
\CombinedResultCheck{cor:monotone}{corollary}\label{cor:monotone}
Under the finite Gaussian model in \cref{sec:finite-model}, suppose
$m\geq4$ and $I\subset[c-L/2,c+L/2]$. Set
$h_1=-\phi_{2,L}-\phi_{4,L}/4$,
$\boldsymbol g=(g_0,g_1,g_3,\ldots,g_m)^T$, and
$f_{\boldsymbol g}=g_0+\sum_{j=1}^mg_j\phi_{j,L}$, where
$g_2=0$, $g_4=a_4-a_2/4$, and $g_j=a_j$ for $j\notin\{2,4\}$.
Then $f_{\boldsymbol a}=f_{\boldsymbol g}+(-a_2)h_1$, with inverse
obtained by reinserting $a_2$ and setting $a_4=g_4+a_2/4$.

For $\theta(x)=\omega_1(x-c+L)$, $w(x)=\cos\theta(x)$, and
$J=[w(x_+),w(x_-)]\subset[-1/\sqrt2,1/\sqrt2]$, define
\begin{equation}\label{eq:monotone-polynomials}
 P_{\boldsymbol g}(w)=\sum_{\substack{1\leq j\leq m\\j\ne2}}jg_jT_j(w),
 \qquad D(w)=1+4w^2-8w^4.
\end{equation}
Since $D\geq1$ on $J$, the threshold is
\begin{equation}\label{eq:monotone-threshold}
 \tau_1(\boldsymbol g)=\max_{w\in J}\frac{-P_{\boldsymbol g}(w)}{D(w)}.
\end{equation}
\cref{thm:finite} applies with scalar coefficient $\gamma=-a_2$.

The threshold is attained at a physical endpoint or a location whose
cosine coordinate is an interior real root in $J$ of
$\frac{\mathrm{d}P_{\boldsymbol g}}{\mathrm{d}w}D-P_{\boldsymbol g}\frac{\mathrm{d}D}{\mathrm{d}w}$.
This stationary polynomial has degree at most $m+3$ when nonzero.
All root multiplicities and reduced degrees are included. If the
stationary polynomial vanishes identically, the quotient
$-P_{\boldsymbol g}/D$ is constant and either endpoint suffices.
A nonzero constant stationary polynomial gives only endpoint candidates.
\end{theoremEnd}

\begin{proofEnd}
\CombinedProofCheck{cor:monotone}
Use the finite Gaussian model in \cref{sec:finite-model} with the rank and closed padding interval in
\cref{cor:monotone}. Substitution of the inverse map proves the curve
identity and invertibility, while the identity
$T_j(w(x))=\cos(j\theta(x))$ gives the derivative formulas
\[
 \frac{\mathrm{d}f_{\boldsymbol g}}{\mathrm{d}x}(x)=\frac{\omega_1}{\sqrt L}P_{\boldsymbol g}(w(x)),
 \qquad \frac{\mathrm{d}h_1}{\mathrm{d}x}(x)=\frac{\omega_1}{\sqrt L}D(w(x)).
\]
The assumed padding confines $\theta(x)$ to $[\pi/4,3\pi/4]$, where
the cosine is strictly decreasing and
$D(w)=1+4w^2(1-2w^2)\geq1$, including at the physical endpoints.
The positive derivative bound therefore allows \cref{thm:finite}
to supply the conditioned law and strictness.
Because the map transports the original Gaussian covariance,
$g_4$ and $a_2$ retain their induced dependence.

Since $D>0$ on compact $J$, the extreme value theorem and Fermat's
theorem locate the maximum of $-P_{\boldsymbol g}/D$ at an endpoint
or an interior stationary point. The quotient rule gives the
stationary polynomial in the statement, whose degree is at most
$m+3$ because $P_{\boldsymbol g}$ and $D$ have degrees at most
$m$ and $4$. If this polynomial vanishes identically, the quotient
is constant by the mean value theorem and either endpoint suffices.
Otherwise it has finitely many distinct roots, each a candidate
when it lies in the interior of $J$, regardless of multiplicity.
This also covers reduced degrees, including a nonzero constant
stationary polynomial with no interior candidates.
\end{proofEnd}

\begin{theoremEnd}[combined main,category=convex]{corollary}[Convexity in the Laplacian expansion]
\CombinedResultCheck{cor:convex}{corollary}\label{cor:convex}
Under the finite Gaussian model in \cref{sec:finite-model}, suppose
$m\geq2$ and $I\subset(c-L,c+L)$. Set $h_2=-\phi_{1,L}$,
$\boldsymbol g=(a_0,a_2,\ldots,a_m)^T$, and
$f_{\boldsymbol g}=a_0+\sum_{j=2}^ma_j\phi_{j,L}$.
Then $f_{\boldsymbol a}=f_{\boldsymbol g}+(-a_1)h_2$, with inverse
obtained by reinserting $a_1$.

For $\theta(x)=\omega_1(x-c+L)$, $w(x)=\cos\theta(x)$, and
$J=[w(x_+),w(x_-)]\subset(-1,1)$, define the polynomial and threshold
\begin{equation}\label{eq:convex-threshold}
 Q_{\boldsymbol g}(w)=\sum_{j=2}^mj^2a_jU_{j-1}(w),
 \qquad \tau_2(\boldsymbol g)=\max_{w\in J}Q_{\boldsymbol g}(w).
\end{equation}
\cref{thm:finite} applies with scalar coefficient $\gamma=-a_1$.

The threshold is attained at a physical endpoint or a location whose
cosine coordinate is an interior real root in $J$ of
$\frac{\mathrm{d}Q_{\boldsymbol g}}{\mathrm{d}w}$. The polynomial $Q_{\boldsymbol g}$
has degree at most $m-1$, so its stationary polynomial has degree at
most $m-2$ when nonzero. All root multiplicities and reduced degrees
are included. If the stationary polynomial vanishes identically,
$Q_{\boldsymbol g}$ is constant and either endpoint suffices.
A nonzero constant stationary polynomial gives only endpoint candidates.
\end{theoremEnd}

\begin{proofEnd}
\CombinedProofCheck{cor:convex}
Use the finite Gaussian model in \cref{sec:finite-model} with the rank and open padding interval in
\cref{cor:convex}. The coordinate map is an invertible signed
permutation. Since $0<\theta(x)<\pi$ on compact $I$,
\[
 \frac{\mathrm{d}^2h_2}{\mathrm{d}x^2}(x)=\frac{\omega_1^2}{\sqrt L}\sin\theta(x)>0,
 \qquad
 \frac{\mathrm{d}^2f_{\boldsymbol g}}{\mathrm{d}x^2}(x)=
 -\frac{\omega_1^2}{\sqrt L}\sin\theta(x)\,Q_{\boldsymbol g}(w(x)).
\]
The second formula uses
$\sin(j\theta(x))=\sin\theta(x)\,U_{j-1}(w(x))$, while compactness
keeps the positive curvature in the first formula bounded away from
zero. Thus \cref{thm:finite} gives the claimed threshold and posterior
under the stated open padding condition.

The extreme value theorem and Fermat's theorem locate the maximum
of $Q_{\boldsymbol g}$ at an endpoint or an interior zero of
$\frac{\mathrm{d}Q_{\boldsymbol g}}{\mathrm{d}w}$. The degree of $Q_{\boldsymbol g}$
is at most $m-1$, so this derivative has degree at most $m-2$ when
nonzero. If it vanishes identically, $Q_{\boldsymbol g}$ is constant
and either endpoint suffices. Otherwise its finitely many distinct
interior roots enter the candidate set regardless of multiplicity,
with none when the derivative is a nonzero constant. The same
argument covers every reduced degree.
\end{proofEnd}

For positivity, separating $a_0$ leaves the spectral curve
$f_{\boldsymbol g}$, and \cref{eq:positive-threshold} is the vertical
shift needed to make its minimum nonnegative. The intercept's
conditional posterior mean can depend on the spectral coefficients,
so reconstruction uses that conditional law even though the prior
coordinates are independent. The monotonicity map makes the same
need to preserve dependence explicit by replacing $a_4$ with
$g_4=a_4-a_2/4$, which is correlated with the separated coefficient
$-a_2$. The convexity map simply sets aside the first mode, but all
three constructions transport the ordinary Gaussian posterior before
conditioning. In the two derivative cases, the intercept stays among
the remaining coefficients and can change its posterior through
correlation, despite contributing neither slope nor curvature.

Although all three constructions use the same padded basis domain
$(c-L,c+L)$, their directions give different sufficient conditions on
the physical interval. The constant positivity direction requires no
additional padding, and $h_2$ has positive curvature throughout the
open basis domain, whereas the slope of the chosen $h_1$ is bounded
away from zero on the central half, including its endpoints. The
conditions in \cref{tab:cases} reflect these properties of the selected
directions, so imposing the central-half condition on convexity would
unnecessarily restrict that finite result.

\begin{table}[htbp]
\centering
\setlength{\tabcolsep}{4pt}
\begin{tabular}{llll}
\toprule
Quantity & Positivity & Monotonicity & Convexity\\
\midrule
Selected order & $r=0$ & $r=1$ & $r=2$\\
Separated coefficient & $a_0$ & $-a_2$ & $-a_1$\\
Direction & $1$ & $-\phi_{2,L}-\phi_{4,L}/4$ & $-\phi_{1,L}$\\
Finite interval & $I\subset(c-L,c+L)$ & $I\subset[c-L/2,c+L/2]$ & $I\subset(c-L,c+L)$\\
Rank condition & $m\geq1$ & $m\geq4$ & $m\geq2$\\
Threshold & $-\min_I f_{\boldsymbol g}$ & $\max_J(-P_{\boldsymbol g}/D)$ & $\max_J Q_{\boldsymbol g}$\\
Polynomial degree bound & $m$ & $m+3$ & $m-2$\\
Limit space & $C(I)$ & $C^1(I)$ & $C^2(I)$\\
Mat\'ern smoothness & $\nu>0$ & $\nu>1$ & $\nu>2$\\
\bottomrule
\end{tabular}
\caption{Positivity, monotonicity, and convexity as alternative
conditioning events in the same spectral model. Here $I$ is the physical
interval and $J$ its image under the decreasing cosine coordinate $w$.
The degree bounds apply to the nonzero polynomials whose roots identify
stationary locations.
The finite rank and
interval conditions are sufficient for the displayed directions. The
function-space limits additionally require \cref{ass:spectral} and the
padding and frequency-cutoff limits in \cref{thm:approximation}.}
\label{tab:cases}
\end{table}

To compute the thresholds in
\cref{eq:positive-threshold,eq:monotone-threshold,eq:convex-threshold},
we compare values at the two interval endpoints and at the interior
stationary locations identified by the polynomial roots in $J$.
For positivity, we evaluate $f_{\boldsymbol g}$ at the corresponding
locations in $I$ and take the negative of the smallest value. For
monotonicity and convexity, we take the largest value of
$-P_{\boldsymbol g}/D$ and $Q_{\boldsymbol g}$, respectively, at the
candidate values of $w$. Every interior real root must be included,
even when the derivative does not change sign there. If the stationary
polynomial vanishes identically, the function being optimized is
constant and an endpoint evaluation suffices.

Nonpositivity, nonincreasing regression, and concavity follow by negating
the observations and prior mean while retaining all covariances, then
applying the corresponding nonnegative-derivative construction and
negating the reconstructed curve and its derivatives. In particular,
concavity makes the slope nonincreasing but allows it to have either
sign, so a curvature restriction alone does not impose monotonicity.
\cref{tab:coefficient-reconstruction} summarizes the inverse maps for
all six individual sign constraints.

\begin{table}[htbp]
\centering
\small
\begin{tabular}{@{}lll@{}}
\toprule
Constraint & Reinserted coefficients & Remaining coefficients\\
\midrule
Positivity & $a_0=\gamma$
  & $a_j=g_j$, $1\leq j\leq m$\\
Negativity & $a_0=-\gamma$
  & $a_j=-g_j$, $1\leq j\leq m$\\
\addlinespace
Nondecreasing & $a_2=-\gamma$, $a_4=g_4-\gamma/4$
  & $a_j=g_j$, $j\notin\{2,4\}$\\
Nonincreasing & $a_2=\gamma$, $a_4=-g_4+\gamma/4$
  & $a_j=-g_j$, $j\notin\{2,4\}$\\
\addlinespace
Convexity & $a_1=-\gamma$
  & $a_j=g_j$, $j\ne1$\\
Concavity & $a_1=\gamma$
  & $a_j=-g_j$, $j\ne1$\\
\bottomrule
\end{tabular}
\caption[Reconstruction for six sign constraints]{Inverse maps to the
original coefficients for six constraints on $I$. Positivity and
negativity mean $f_{\boldsymbol a}\geq0$ and $f_{\boldsymbol a}\leq0$.
The negative, nonincreasing, and concave cases use
$(\boldsymbol g,\gamma)$ from the Gaussian calculation for
$-f_{\boldsymbol a}$ with negated observations and prior mean.
The other cases use $f_{\boldsymbol a}$. Indices range from $0$ to $m$
unless stated otherwise, and $g_j$ retains its original coefficient
index. Each pair uses the rank and padding conditions of its
corresponding case in \cref{cor:positive,cor:monotone,cor:convex}.}
\label{tab:coefficient-reconstruction}
\end{table}

Several restrictions can describe different aspects of the same
response. A nonnegative, nondecreasing, concave curve, for example,
has a nonnegative level and slope, with the slope nonincreasing across
the interval. We express such intersections by selecting derivative
orders and a sign at each order. For a nonempty index set
$A\subseteq\{0,1,2\}$ and a sign choice
$\varepsilon:A\to\{-1,1\}$, write $\varepsilon_r=\varepsilon(r)$ and define
\begin{equation}\label{eq:joint-event}
 \mathcal C_A^\varepsilon=\left\{f\in C^{\max A}(I):
       \varepsilon_r\frac{\mathrm{d}^rf}{\mathrm{d}x^r}(x)\geq0
       \text{ for every }r\in A\text{ and }x\in I\right\}.
\end{equation}
At order zero, the two signs give nonnegativity and nonpositivity.
At orders one and two, they give nondecreasingness and nonincreasingness,
and convexity and concavity, respectively. Taking $\varepsilon_r=1$
at every selected order gives positivity, nondecreasingness, and
convexity, their three pairwise combinations, and their triple
combination. Allowing either sign gives six individual cases and
20 joint combinations without opposing signs. Imposing both signs
at the same order instead forces the corresponding derivative to
vanish throughout $I$, an event of zero probability under the finite
Gaussian model.

To retain the scalar conditioning construction, we need one direction
$h$ in the retained span whose selected signed derivatives are all
uniformly positive on $I$. Increasing its coefficient then increases
every signed derivative involved in the restriction. Each order and
location imposes a lower bound on this same coefficient, so their
largest bound enforces the entire intersection. The direction
condition must be verified for the chosen combination, rank, and
interval. The following proposition gives the resulting Gaussian
conditioning law.

\begin{theoremEnd}[combined main,category=joint]{proposition}[Joint sign constraints]
\CombinedResultCheck{prop:joint}{proposition}\label{prop:joint}
Under the finite Gaussian model in \cref{sec:finite-model}, let
$\boldsymbol a=(a_0,\ldots,a_m)^T$ denote the coefficient vector and fix
a nonempty $A\subseteq\{0,1,2\}$ and signs
$\varepsilon:A\to\{-1,1\}$, with event $\mathcal C_A^\varepsilon$ defined in
\cref{eq:joint-event}. Suppose an invertible linear change of coordinates
$\boldsymbol a\mapsto(\boldsymbol g,\gamma)\in\R^m\times\R$
gives $f_{\boldsymbol a}=f_{\boldsymbol g}+\gamma h$.
Here $h$ is a fixed function in the retained span, shared by all selected
orders, $\gamma$ is its scalar coefficient, and $f_{\boldsymbol g}$
depends linearly on the remaining coordinates $\boldsymbol g$.
Assume $h,f_{\boldsymbol g}\in C^{\max A}(I)$ with
$\min_{x\in I}\varepsilon_r\frac{\mathrm{d}^rh}{\mathrm{d}x^r}(x)>0$
for every $r\in A$. Define
\begin{equation}\label{eq:joint-threshold}
 \eta_r(x;\boldsymbol g)=
       \frac{-\varepsilon_r\frac{\mathrm{d}^rf_{\boldsymbol g}}{\mathrm{d}x^r}(x)}
            {\varepsilon_r\frac{\mathrm{d}^rh}{\mathrm{d}x^r}(x)},
 \qquad
 \tau_A^\varepsilon(\boldsymbol g)=
       \max_{r\in A}\max_{x\in I}\eta_r(x;\boldsymbol g).
\end{equation}
This finite, attained threshold satisfies
$f_{\boldsymbol a}\in\mathcal C_A^\varepsilon$ exactly when
$\gamma\geq\tau_A^\varepsilon(\boldsymbol g)$.
The event has positive ordinary posterior probability and a null
coefficient boundary. Its marginal density and reconstruction are
\cref{eq:marginal,eq:reconstruction}, with $\tau_r,\mathcal C_r$
replaced by $\tau_A^\varepsilon,\mathcal C_A^\varepsilon$ and the
Gaussian parameters in \cref{eq:conditional} computed under this map.
The conditioned law satisfies
$\min_{x\in I}\varepsilon_r\frac{\mathrm{d}^rf_{\boldsymbol a}}{\mathrm{d}x^r}(x)>0$
almost surely for every $r\in A$. The envelope, strong log-concavity,
and conservative threshold-error conclusions extend under their
respective hypotheses.
\end{theoremEnd}

\begin{proofEnd}
\CombinedProofCheck{prop:joint}
Fix the coefficient map and signed direction satisfying the hypotheses
of \cref{prop:joint}.
Dividing each signed derivative inequality by its positive direction
derivative gives $\gamma\geq\eta_r(x;\boldsymbol g)$, so their intersection
gives \cref{eq:joint-threshold}. Compactness of $I$, finiteness
of $A$, and the denominator bounds make the maximum finite and attained.
The Gaussian conditioning argument in the proof of \cref{thm:finite} then gives
positive probability, the marginal density, and reconstruction.

Above the threshold, every selected signed derivative has a strictly
positive minimum. At equality, an attained maximizing pair gives a zero
derivative, and perturbing $\gamma$ crosses the event boundary. Thus the
boundary is the threshold graph, which has zero conditional Gaussian
probability for every $\boldsymbol g$, establishing almost-sure strictness. Each
$\eta_r(x;\boldsymbol g)$ is linear in $\boldsymbol g$, so the envelope
and strong log-concavity arguments apply with the compact index set
$A\times I$, and \cref{prop:error} applies to the same scalar truncation.
\end{proofEnd}

The joint restrictions can also reduce the number of continuum
searches. For the nonnegative, nondecreasing, concave example, take
$A=\{0,1,2\}$ with $\varepsilon_0=\varepsilon_1=1$ and
$\varepsilon_2=-1$. Concavity makes the slope smallest at $x_+$,
and nondecreasingness makes the response smallest at $x_-$.
Under the common map in \cref{prop:joint}, the threshold therefore becomes
\[
 \tau_A^\varepsilon(\boldsymbol g)
 =\max\left\{
       \max_{x\in I}\eta_2(x;\boldsymbol g),
       \eta_1(x_+;\boldsymbol g),\eta_0(x_-;\boldsymbol g)
       \right\}.
\]
Thus this combination requires one continuum curvature bound and
two endpoint bounds. All three use the same $h$ and $\boldsymbol g$,
whose choice determines the quotients in \cref{eq:joint-threshold}.
The supplementary material gives the HSGP coefficient map, conditions
on its direction coefficients, and the reductions for all sign choices,
with a table separating continuum and endpoint bounds for the 20
joint cases.

\section{Geometry, numerical error, and approximation}\label{sec:theory}

At a fixed location $x\in I$, the ratio in \cref{eq:threshold} is
the lower bound on $\gamma$ required to satisfy the sign restriction
at that location. Because $f_{\boldsymbol g}$ depends linearly on
$\boldsymbol g$ and $h_r$ is fixed, this bound is a linear function
$\boldsymbol\beta_r(x)^T\boldsymbol g$. Here
$\boldsymbol\beta_r(x)\in\R^m$ is the vector of coefficient
multipliers, with its entries ordered as those of $\boldsymbol g$.
For positivity, for example, $h_0=1$ and
$\boldsymbol g=(a_1,\ldots,a_m)^T$, so
$\boldsymbol\beta_0(x)=(-\phi_{1,L}(x),\ldots,-\phi_{m,L}(x))^T$.
The continuum threshold is therefore
\[
 \tau_r(\boldsymbol g)
   =\max_{x\in I}\boldsymbol\beta_r(x)^T\boldsymbol g.
\]

The continuity assumptions and positive derivative bound in
\cref{thm:finite} make $\boldsymbol\beta_r$ continuous and bounded
on $I$, so the maximum representation makes $\tau_r$ convex,
positively homogeneous, and globally Lipschitz.
The threshold enters the log posterior through the normal survival
weight, so its derivatives are needed to compute the gradients used
in posterior sampling. Danskin's theorem \citep{danskin1966maxmin}
expresses these derivatives through the coefficient vectors at
maximizing locations. When the maximizing location $x_\star$ is
unique, the threshold has gradient $\boldsymbol\beta_r(x_\star)$.
Several maximizing locations also permit a gradient when their
coefficient vectors agree, whereas different vectors produce kinks.
The supplementary material gives the threshold derivatives and the
resulting log-posterior gradient formulas.
For a joint restriction in \cref{prop:joint},
the vectors are formed under the common coefficient map and the
maximum ranges over pairs $(r,x)\in A\times I$. A change in the
maximizing derivative order can therefore produce a kink as well
as a change in its location.

Let $\boldsymbol B\boldsymbol B^T=\boldsymbol V_g$ with
$\boldsymbol B$ invertible, and write
$\boldsymbol g(\boldsymbol z)=\boldsymbol m_g+\boldsymbol B\boldsymbol z$.
The exact whitened log density, up to a constant, is
\begin{equation}\label{eq:whitened-target}
 -\tfrac12\|\boldsymbol z\|^2+
 \log\Phi^{\mathrm c}\left(\frac{\tau_r(\boldsymbol g(\boldsymbol z))
                         -\mu(\boldsymbol g(\boldsymbol z))}{s}\right).
\end{equation}
The supplementary material establishes that this density is
$1$-strongly log-concave, giving a unique mode and finite moments of
every polynomial order.

The continuum maximum generally requires numerical evaluation, and
an underestimated threshold would admit curves that violate the
intended sign restriction. We therefore use a conservative upper
threshold $\widetilde\tau_r\geq\tau_r$, so that imposing
$\gamma\geq\widetilde\tau_r(\boldsymbol g)$ preserves the restriction.
The cost is that admissible coefficient values between the exact and
upper thresholds are excluded from the posterior.
The upper threshold may not be
convex, so the log-concavity conclusion need not carry over to this
numerical posterior. Its difference from the exact law can instead
be quantified through the probability mass of the excluded
coefficients, as the following proposition shows under exact Gaussian
operations.

\begin{theoremEnd}[combined main,category=error]{proposition}[Posterior error from a conservative continuum threshold]
\CombinedResultCheck{prop:error}{proposition}\label{prop:error}
Under \cref{thm:finite}, let $\widetilde\tau_r$ be measurable and satisfy
$\tau_r\leq\widetilde\tau_r\leq\tau_r+\delta$ almost everywhere
under the ordinary posterior of $\boldsymbol g$, for a fixed $\delta\geq0$.
Use the same upper threshold in marginal weighting and reconstruction.
Let $P_{\boldsymbol y}$ be the ordinary coefficient posterior, and use
$\mathcal C_r$ also for its coefficient preimage. Define
$\widetilde{\mathcal C}_r=\{\gamma\geq\widetilde\tau_r(\boldsymbol g)\}$.
Both conditioned laws exist and, with $\TV(P,Q)=\sup_A|P(A)-Q(A)|$,
\begin{equation}\label{eq:tv}
 \begin{split}
 \TV\bigl(P_{\boldsymbol y}(\cdot\mid\widetilde{\mathcal C}_r),
          P_{\boldsymbol y}(\cdot\mid\mathcal C_r)\bigr)
 &=1-\frac{P_{\boldsymbol y}(\widetilde{\mathcal C}_r)}
                 {P_{\boldsymbol y}(\mathcal C_r)}
 \leq\min\left(1,\frac{\delta}
 {s\sqrt{2\pi}P_{\boldsymbol y}(\mathcal C_r)}\right).
 \end{split}
\end{equation}
Every measurable summary of the reconstructed coefficients inherits
this bound. The identity holds for any inner event of positive probability.
\end{theoremEnd}
\begin{proofEnd}
\CombinedProofCheck{prop:error}
Fix the ordinary Gaussian law, selected order, and almost-everywhere
error condition of \cref{prop:error}, modifying the numerical threshold
on the exceptional null set if necessary. Finite thresholds and $s>0$
give positive conditional tails, so both conditioning events have
positive probability. To compare the resulting laws, put
$A=\mathcal C_r$ and $B=\widetilde{\mathcal C}_r$. If the omitted
mass is zero, both conditioned laws agree. Otherwise the law on $A$
is a mixture of the laws on $B$ and $A\setminus B$, with weights
$P_{\boldsymbol y}(B)/P_{\boldsymbol y}(A)$ and its complement.
The components have disjoint supports, so total variation equals the
second weight, establishing the identity for any positive-probability
inner event.

For the threshold events considered here, conditional
Gaussian integration identifies the omitted mass as
\[
 \E\left[\Phi((\widetilde\tau_r(\boldsymbol g)-\mu(\boldsymbol g))/s)
          -\Phi((\tau_r(\boldsymbol g)-\mu(\boldsymbol g))/s)
          \mid\boldsymbol y\right].
\]
Since the conditional density is at most $1/(s\sqrt{2\pi})$ and the
excluded interval has length at most $\delta$, dividing this mass
bound by the probability of the exact event proves \cref{eq:tv}.
The invertible coefficient map preserves total variation, while
measurable summaries cannot increase it, proving the remaining assertion.
\end{proofEnd}

The conditioning probability in \cref{eq:tv} explains why a rare
shape event is more sensitive to a given absolute threshold error.
The bound requires the threshold
guarantee to hold globally or almost everywhere under the ordinary
posterior and quantifies the effect of threshold arithmetic within
the finite model, leaving basis approximation and finite-chain error
as separate quantities.

\subsection{Convergence to the conditioned underlying GP}

We next consider the error introduced by replacing the underlying GP
with a finite HSGP expansion, and how this error depends on padding
and rank.
Increasing $L$ separates the artificial boundary from the physical
interval $I$, whereas $m/L$ controls the frequency cutoff, so increasing
padding at fixed rank can remove variation at short scales. The two
choices must therefore be considered together and for the derivative
used in inference as well as for the response
\citep{riutort2023practicalhsgp}.

Because the conditioning event depends on the response or its selected
derivative between grid points, convergence of covariance blocks on
a finite grid is insufficient for comparing the conditioned laws.
We instead work on the entire physical interval in
$C^r(I)$ with its Borel sigma-field and norm
\[
 \|f\|_{C^r}=\sum_{j=0}^r\sup_{x\in I}
             \left|\frac{\mathrm{d}^jf}{\mathrm{d}x^j}(x)\right|,
 \qquad \frac{\mathrm{d}^0f}{\mathrm{d}x^0}=f.
\]

\begin{theoremEnd}{assumption}[Common underlying and finite Gaussian models]
\CombinedResultCheck{ass:underlying}{assumption}\label{ass:underlying}
Fix $I\subset[c-R,c+R]$ with $R>0$ and the observation model and
intercept law specified in \cref{sec:finite-model}. Let $F=a_0+W$, where $W$ is
a centered stationary GP on $\R$, independent of $a_0$, with covariance
function $k$ and spectral density $S_k$ as defined in
\cref{sec:finite-model}. Use that same spectrum and intercept in the finite models in
\cref{eq:finite-model}, varying only $m$ and $L$ while keeping the
center, interval, spectrum, intercept law, observation locations,
positive-definite noise covariance, and data identical in all models.
\end{theoremEnd}

\begin{theoremEnd}{assumption}[Spectral regularity for the selected order]
\CombinedResultCheck{ass:spectral}{assumption}\label{ass:spectral}
Fix $r\in\{0,1,2\}$. The spectrum $S_k$ is continuous, even, and
strictly positive. There is a regularity parameter $\zeta\in(0,1]$
such that the two functions
\[
 \omega\longmapsto S_k(\omega),
 \qquad \omega\longmapsto\omega^{2r+2\zeta}S_k(\omega)
\]
are locally absolutely continuous on $[0,\infty)$, and each function
and its derivative belongs to $L^1(0,\infty)$.
The weighted spectrum $\omega^{2r+2\zeta}S_k(\omega)$ tends to zero
as $\omega\downarrow0$, and we use this limiting value at the origin.
The frequency exponent is
$2\zeta$ for positivity, $2+2\zeta$ for monotonicity, and
$4+2\zeta$ for convexity.
\end{theoremEnd}

Under the Fourier representation of the stationary covariance,
differentiating $r$ times with respect to each input multiplies
the spectrum by $\omega^{2r}$. The additional factor $\omega^{2\zeta}$
allows us to bound the variance of the difference between the $r$th
derivative values at $x$ and $x'$ by a constant multiple of
$|x-x'|^{2\zeta}$. With $\zeta$ held fixed in the approximation limit,
this increment bound supplies the additional regularity needed for
continuity and tightness in $C^r(I)$.

Let $\Pi_{m,L,\boldsymbol y}^{\mathcal C_r}$ and
$\Pi_{\boldsymbol y}^{\mathcal C_r}$ denote the finite and underlying
posterior laws obtained by weighting their priors by the common
Gaussian likelihood and the indicator of \cref{eq:shape-event}.
For laws with finite second norm moments, define
\[
 W_2(P,Q)=\inf_{\lambda}
 \left(\int\|f-g\|_{C^r}^2\,\lambda(df,dg)\right)^{1/2},
\]
where $\lambda$ ranges over couplings with marginals $P,Q$.

Convergence in $W_2$ means that the finite-model curve $F_{m,L}$ and
the underlying curve $F$, with their respective posterior distributions
conditional on the data and $\mathcal C_r$, can be coupled so that
\[
 \E\left[
 \sup_{x\in I}
 \left|
 \frac{\mathrm{d}^jF_{m,L}}{\mathrm{d}x^j}(x)
 -
 \frac{\mathrm{d}^jF}{\mathrm{d}x^j}(x)
 \right|^2
 \right]\longrightarrow0,
 \qquad j=0,\ldots,r,
\]
as $L\to\infty$ and $m/L\to\infty$. Under this coupling, the
expected squared maximum discrepancy over $I$ vanishes for the
response and each derivative through order $r$.

\begin{theoremEnd}[combined main,category=approximation]{theorem}[Convergence of shape-constrained HSGP posteriors]
\CombinedResultCheck{thm:approximation}{theorem}\label{thm:approximation}
Fix $r\in\{0,1,2\}$ under \cref{ass:underlying,ass:spectral}, and
consider finite models with $L\geq2R$ and $m\geq1$ for $r=0$,
$m\geq4$ for $r=1$, or $m\geq2$ for $r=2$.
The underlying GP has a $C^r(I)$ version, and the finite and underlying
shape events have positive probability and null boundaries in that space.
Consequently, the posteriors conditional on the data and $\mathcal C_r$
exist, with finite moments of every polynomial order in $\|\cdot\|_{C^r}$.
As $L\to\infty$ and $m/L\to\infty$,
\begin{equation}\label{eq:wasserstein}
 W_2\bigl(\Pi_{m,L,\boldsymbol y}^{\mathcal C_r},
          \Pi_{\boldsymbol y}^{\mathcal C_r}\bigr)\longrightarrow0.
\end{equation}
This convergence also holds for the posterior summaries: means of
$\frac{\mathrm{d}^jf}{\mathrm{d}x^j}$ converge uniformly on $I$ for
every $0\leq j\leq r$, and covariances between
$\frac{\mathrm{d}^{r_1}f}{\mathrm{d}x^{r_1}}(x)$ and $\frac{\mathrm{d}^{r_2}f}{\mathrm{d}x^{r_2}}(x')$
converge uniformly on $I\times I$ for every $0\leq r_1,r_2\leq r$.
Under the conditioned underlying law,
$\min_I\frac{\mathrm{d}^rf}{\mathrm{d}x^r}>0$ almost surely.
\end{theoremEnd}
\begin{proofEnd}
\CombinedProofCheck{thm:approximation}
Fix $r$ and any sequence of models satisfying the conditions and limits
in \cref{thm:approximation}.
Write $F_{m,L}$ for a draw from the finite prior and
$k_{m,L}(x,x')=\sum_{j=1}^mS_k(\omega_j)\phi_{j,L}(x)\phi_{j,L}(x')$
for the covariance of its spectral component.
We first establish weak convergence of the priors in $C^r(I)$ and
uniform norm moments. Positive shape probability and a boundary of
prior probability zero will then allow passage to the conditioned
posteriors, whose moment bounds give Wasserstein convergence and
convergence of the stated summaries.

\paragraph{Covariance and increment bounds.}
Put $b=2r+2\zeta$. The two spectral weights in
\cref{ass:spectral} also control the intermediate orders needed for
covariance derivatives. For each integer $0<d\leq2r$, differentiating
$\omega^dS_k(\omega)$ on $[0,1]$ gives integrable terms because $S_k$
is bounded there and its derivative is integrable. On $[1,\infty)$,
write the same function as $\omega^{d-b}(\omega^bS_k(\omega))$.
The multiplier and its derivative are bounded, so the product and its
derivative are integrable. Thus, for $d=0,\ldots,2r$ and $d=b$,
the weighted spectra are locally absolutely continuous and have
themselves and their derivatives in $L^1(0,\infty)$, with the cases
$d=0$ and $d=b$ supplied directly by the assumption. Their finite
values at zero and integrable derivatives also give boundedness.
The derivative-covariance argument of
\citet[Theorem 2 and its proof]{differentialensembles2026} then gives, for
$0\leq r_1,r_2\leq r$, with $d=r_1+r_2$:
\begin{equation}\label{eq:covariance-bound}
 \sup_{x,x'\in I}\left|
 \frac{\partial^{r_1+r_2}(k_{m,L}(x,x')-k(x-x'))}
      {\partial x^{r_1}\partial(x')^{r_2}}\right|
 \leq\frac{b_d}{L}+\frac1\pi\int_{\pi m/(2L)}^\infty
                       \omega^dS_k(\omega)\,d\omega,
\end{equation}
where $b_d<\infty$ is independent of $m$ and $L$.
The cited theorem states stronger conditions on classical derivatives,
but its proof bounds frequency-cell errors by integrals of absolute
derivatives and the reflected term by consecutive frequency differences.
Both steps require only the absolute continuity, integrability, and
boundedness verified above. In \cref{eq:covariance-bound}, the first
term vanishes as the padded boundary recedes, while the second vanishes
as the frequency cutoff increases. The common intercept adds no
covariance error.

For the frequency spacing $\Delta=\pi/(2L)$, with $\omega_j=j\Delta$,
the same frequency-cell estimate bounds the spectral sums by
\[
 \Delta\sum_{j=1}^m(j\Delta)^dS_k(j\Delta)
 \leq\|\omega^dS_k(\omega)\|_1+
 \Delta\left\|\frac{\mathrm{d}(\omega^dS_k(\omega))}{\mathrm{d}\omega}\right\|_1,
 \qquad d\in\{2j:0\leq j\leq r\}\cup\{b\}.
\]
Here the norms are over $[0,\infty)$. The right-hand side is uniformly
bounded because $L\geq2R$ gives $\Delta\leq\pi/(4R)$.
Taking $d=0,2,\ldots,2r$ therefore bounds the derivative variances at
$x_-$, with the intercept adding its fixed variance at order zero.

To control increments of the highest derivative, we use this estimate
with $d=b=2r+2\zeta$. Each differentiated basis function is a sine or
cosine with amplitude $L^{-1/2}\omega_j^r$, so
$|\sin u-\sin v|\leq2^{1-\zeta}|u-v|^\zeta$ and the corresponding
cosine inequality give
\[
 \left|\frac{\mathrm{d}^r\phi_{j,L}}{\mathrm{d}x^r}(x)-
        \frac{\mathrm{d}^r\phi_{j,L}}{\mathrm{d}x^r}(x')\right|
 \leq\frac{2^{1-\zeta}}{\sqrt L}
       \omega_j^{r+\zeta}|x-x'|^\zeta.
\]
By independence of the Gaussian coefficients, the variance of the
$r$th derivative increment is the sum of the squared basis increments
above weighted by $S_k(\omega_j)$. The intercept cancels in increments when $r=0$ and
contributes no derivative when $r\geq1$. Consequently,
\[
 \Var\left(\frac{\mathrm{d}^rF_{m,L}}{\mathrm{d}x^r}(x)-
           \frac{\mathrm{d}^rF_{m,L}}{\mathrm{d}x^r}(x')\right)
 \leq C_\zeta|x-x'|^{2\zeta}
       \Delta\sum_{j=1}^m\omega_j^{2r+2\zeta}S_k(\omega_j),
\]
where $C_\zeta=2^{3-2\zeta}/\pi$ follows from $1/L=2\Delta/\pi$.
The frequency-cell estimate at $d=b$ bounds the spectral sum on the
right uniformly in $m,L$, giving the increment bound
$A|x-x'|^{2\zeta}$ with $A$ independent of rank and padding.
Fourier inversion gives the same bound for the underlying mean-square
$r$th derivative, after increasing $A$ if needed. Gaussian moments
then verify the Kolmogorov continuity criterion
\citep{kallenberg2021foundations} with any exponent $p$ satisfying
$p\zeta>1$.
For $r=0$, this gives a continuous version of $W$. For $r\geq1$,
integrating the continuous highest derivative, with the lower
mean-square derivative values at $x_-$ as integration constants,
recovers a version of $W$ by the mean-square fundamental theorem of
calculus. Adding the intercept therefore gives a $C^r(I)$ version of $F$.

\paragraph{Tightness and norm moments.}
We next control both the size and the continuity of the sample paths
uniformly over the finite models. These bounds establish tightness
in $C^r(I)$ and provide the norm moments needed later for Wasserstein
convergence and convergence of posterior summaries.
Let $H$ denote either $\frac{\mathrm{d}^rF_{m,L}}{\mathrm{d}x^r}$ or
$\frac{\mathrm{d}^rF}{\mathrm{d}x^r}$, and choose
$p>\max(4,1/\zeta)$ and a H\"older exponent $0<\kappa<\zeta-1/p$.
For $j\geq0$, let $M_j$ be the largest absolute increment of $H$
between neighboring points on the mesh dividing $I$ into $2^j$
equal cells. The Gaussian increment bound gives
$\E[M_j^p]^{1/p}\leq C_p2^{-(\zeta-1/p)j}$ uniformly over the models,
because the $p$th power of the maximum is bounded by the sum of the
$p$th powers of the $2^j$ increments.
Approximating arbitrary $x,x'\in I$ by points on successively finer
meshes expresses $H(x)-H(x')$ through adjacent increments.
This dyadic chaining argument bounds the H\"older seminorm by
\[
 [H]_\kappa
 =\sup_{\substack{x,x'\in I\\x\ne x'}}
      \frac{|H(x)-H(x')|}{|x-x'|^\kappa}
 \leq C_{I,\kappa}\sum_{j=0}^{\infty}2^{\kappa j}M_j.
\]
Here $C_{I,\kappa}$ depends only on $I$ and $\kappa$.
Minkowski's inequality therefore bounds $\E[[H]_\kappa^p]^{1/p}$
by a constant times $\sum_{j\geq0}2^{-(\zeta-1/p-\kappa)j}$.
Our choice of $\kappa$ makes this series converge, giving a uniform
$p$th moment bound for the seminorm.

The seminorm and the value $H(x_-)$ together bound the highest
derivative everywhere on $I$. When $r\geq1$, repeated integration,
starting from the lower derivative values at $x_-$, then bounds
every lower derivative and the curve itself. For $r=0$, the first
bound already controls the curve. The uniformly bounded Gaussian
moments of these endpoint values consequently give
\begin{equation}\label{eq:moments}
 \sup_{m,L}\E[\|F_{m,L}\|_{C^r}^p]+\E[\|F\|_{C^r}^p]<\infty
\end{equation}
over the permitted ranks and padding. Since $p$ can be arbitrarily
large, every polynomial norm moment is uniformly bounded.

Bounds on derivative values of orders $0,\ldots,r$ at $x_-$ and on the highest
derivative's H\"older seminorm define relatively compact subsets of
$C^r(I)$ by Arzel\`a--Ascoli. The moment bounds and Markov's inequality
allow these bounds to be chosen so that the finite priors assign
arbitrarily high probability to such a set, uniformly over the models.
This proves tightness. To identify the limit, the common mean and
\cref{eq:covariance-bound} give convergence of the joint Gaussian laws
of every finite collection of derivative values. Such evaluations on
a countable dense subset of $I$ determine a $C^r(I)$ law, so every
subsequential limit is the underlying prior. The finite priors
therefore converge weakly to that law.

\paragraph{Positive shape probability and a null boundary.}
The boundary of $\mathcal C_r$ in $C^r(I)$ consists exactly of curves
with $\min_I\frac{\mathrm{d}^rf}{\mathrm{d}x^r}=0$. Indeed, this minimum is continuous
in the $C^r$ norm, and adding or subtracting a small multiple of
$(x-c)^r/r!$ moves any such curve to either side of the event.
We therefore need positive probability of nonnegativity and zero
probability of equality. For each order, these follow by separating
one Gaussian coordinate whose contribution to the selected derivative
is positive throughout $I$.

For $r=0$, the independent intercept gives
\[
 \min_{x\in I}F(x)=a_0+\min_{x\in I}W(x).
\]
Conditional on $W$, nonnegativity therefore imposes a finite lower
threshold on the nondegenerate Gaussian $a_0$, giving positive
conditional probability and zero conditional probability of equality.

For $r\in\{1,2\}$, the intercept contributes no derivative. We instead
seek a decomposition of $W$ into a Gaussian coefficient multiplying
a deterministic function whose $r$th derivative is uniformly positive,
together with an independent Gaussian process. This reduces the shape
restriction to a scalar threshold, as in the finite model.
The reproducing kernel Hilbert space (RKHS) $\mathcal H$ of $W$ in
$C^r(I)$ identifies the deterministic directions that can be separated
in this way.

To construct a candidate direction, choose
$0<\omega_-<\omega_+<\pi/(2R)$ and set
\[
 v_r(x)=\int_{\omega_-}^{\omega_+}S_k(\omega)
             \cos(\omega(x-c)-r\pi/2)\,d\omega.
\]
Taking $r$ derivatives removes the phase shift, leaving the cosine
factor $\cos(\omega(x-c))$. Because $|x-c|\leq R$ and
$\omega_+R<\pi/2$, this factor is uniformly positive on the selected
frequency band and throughout $I$. Thus
\[
 \frac{\mathrm{d}^rv_r}{\mathrm{d}x^r}(x)\geq
 \cos(\omega_+R)\int_{\omega_-}^{\omega_+}\omega^rS_k(\omega)\,d\omega>0.
\]

We next verify that this direction belongs to $\mathcal H$, which
justifies separating a Gaussian coefficient along it.
The Fourier expression for the full-line RKHS norm gives squared norm
$\pi\int_{\omega_-}^{\omega_+}S_k(\omega)\,d\omega<\infty$ for $v_r$,
and restricting the function to $I$ cannot increase that norm.
The RKHS obtained from the covariance on $I$ agrees with the RKHS
of $W$ as a $C^r(I)$-valued Gaussian random element by
\citet[Lemma 8.1]{vdv2008rkhs}, applied to the continuous, injective
embedding into $C(I)$. The norm calculation therefore establishes
$v_r\in\mathcal H$ in the function space used in this proof.

Since $v_r$ is nonzero, we can normalize it by its RKHS norm.
The Gaussian RKHS isometry \citep[Lemma 2.3]{vdv2008rkhs} then supplies
a standard normal variable $\xi$, jointly Gaussian with $W$, such that
$\E[\xi W(x)]=v_r(x)/\|v_r\|_{\mathcal H}$ for every $x\in I$.
Define the remaining process $V$ by
\[
 W=\xi\frac{v_r}{\|v_r\|_{\mathcal H}}+V.
\]
Subtracting this component preserves $C^r(I)$ paths, and
$\Var(\xi)=1$ gives $\Cov(\xi,V(x))=0$ for every $x\in I$.
Joint Gaussianity therefore makes $\xi$ independent of the whole
process $V$.

Conditional on $V$, only the scalar coefficient $\xi$ remains random
in the derivative constraint. Dividing each derivative inequality by
the uniformly positive derivative of $v_r$ gives
\[
 \{F\in\mathcal C_r\}=
 \left\{\xi\geq\|v_r\|_{\mathcal H}
       \max_{x\in I}\frac{-\frac{\mathrm{d}^rV}{\mathrm{d}x^r}(x)}
                              {\frac{\mathrm{d}^rv_r}{\mathrm{d}x^r}(x)}\right\}.
\]
For each path of $V$ the threshold is finite, so the conditional normal
tail has positive probability and no atom at equality. Equality
is precisely the zero-minimum event identified above, so the underlying
shape event has positive probability and a null boundary at all three
orders. The finite laws have the same properties by
\cref{cor:positive,cor:monotone,cor:convex}, applied with
likelihood one. In both cases, conditioning on nonnegativity therefore
gives a strictly positive minimum selected derivative almost surely.

\paragraph{Likelihood and shape conditioning.}
Write $\mathcal L_{\boldsymbol y}(f)$ for the Gaussian density at
$\boldsymbol y$ with mean $(f(x_1),\ldots,f(x_n))^T$ and covariance
$\boldsymbol\Sigma$, taking $\mathcal L_{\boldsymbol y}(f)=1$ when
$n=0$. This likelihood is bounded, positive, and continuous on
$C^r(I)$. For any bounded continuous test function
$\psi:C^r(I)\to\R$, the map
\[
 f\longmapsto\psi(f)\mathcal L_{\boldsymbol y}(f)
                    \boldsymbol1_{\mathcal C_r}(f)
\]
is bounded and continuous outside $\partial\mathcal C_r$, where
$\boldsymbol1_{\mathcal C_r}(f)$ is one if $f\in\mathcal C_r$ and
zero otherwise. The preceding argument shows that this boundary
has zero probability under the underlying prior, so the Portmanteau theorem
\citep[Section 2]{billingsley1999convergence}
gives convergence of the integrals of this map under the finite priors
to its integral under the underlying prior.
Taking $\psi\equiv1$ shows that the normalizing constants converge
to a positive limit. Dividing by them gives convergence of posterior
expectations of every bounded continuous $\psi$, which is conditioned
weak convergence. The posterior densities relative to the priors are
uniformly bounded along the sequence by the likelihood supremum
divided by these constants, so
\cref{eq:moments} also holds for the conditioned laws at every fixed
polynomial order. Likelihood weighting preserves the zero boundary
probability, giving the strictness assertion.

\paragraph{Wasserstein convergence and posterior summaries.}
We now strengthen conditioned weak convergence to convergence in
Wasserstein distance. Index the chosen approximation sequence by
$\ell$, with ranks $m_\ell$ and padding $L_\ell$. In this paragraph,
write $F_\ell$ and $F$ for draws from the finite and underlying
conditioned posteriors, respectively. Since $C^r(I)$ is separable
and complete, the Skorokhod representation theorem
\citep[Theorem 6.7]{billingsley1999convergence} allows these draws
to be constructed on a common probability space, preserving their
posterior laws, so that $\|F_\ell-F\|_{C^r}\to0$ almost surely.
The uniform fourth-moment bounds established above make the squared
norm differences uniformly integrable, so their expectations also
converge to zero. Wasserstein distance takes the infimum over all
couplings, and this particular coupling therefore gives
\[
 W_2^2\bigl(\Pi_{m_\ell,L_\ell,\boldsymbol y}^{\mathcal C_r},
            \Pi_{\boldsymbol y}^{\mathcal C_r}\bigr)
 \leq \E[\|F_\ell-F\|_{C^r}^2]\longrightarrow0,
\]
proving \cref{eq:wasserstein}.

The same coupling controls posterior summaries uniformly in their
input locations because the $C^r$ norm bounds every derivative through
order $r$ over the whole interval. For each such order, the uniform
difference of derivative means is at most
$\E[\|F_\ell-F\|_{C^r}]$, which tends to zero by Cauchy--Schwarz.
For any two derivative orders through $r$, expand the difference of
their products by changing one factor at a time. The resulting
difference of mixed second moments is bounded, uniformly on
$I\times I$, by
\[
 \E[\|F_\ell-F\|_{C^r}(\|F_\ell\|_{C^r}+\|F\|_{C^r})].
\]
Cauchy--Schwarz bounds this expectation by the product of
$\E[\|F_\ell-F\|_{C^r}^2]^{1/2}$, which tends to zero, and
$\E[(\|F_\ell\|_{C^r}+\|F\|_{C^r})^2]^{1/2}$, which remains bounded.
Thus the mixed second moments converge uniformly. Subtracting the
uniformly convergent products of means proves covariance convergence.
Since the sequence was arbitrary, all these conclusions hold in the
full stated limiting regime.
\end{proofEnd}

The spectral weight controls the highest derivative in the
function-space norm, so positivity needs only a continuous process,
whereas monotonicity and convexity require $C^1$ and $C^2$ regularity,
respectively. The proof adapts the derivative-covariance argument of
\citet[Theorem 2 and its proof]{differentialensembles2026} to the
weighted-spectral conditions above, then uses increment moments to
obtain convergence in that space and passes through likelihood
and event conditioning. This last step requires positive event
probability and a null boundary, which the independent random intercept
supplies for positivity and a positive reproducing kernel Hilbert space
(RKHS) derivative direction supplies for the two derivative constraints.

Squared exponential spectra satisfy all three sets of conditions. A Mat\'ern
spectrum is proportional to
$(\lambda^2+\omega^2)^{-(\nu+1/2)}$ with $\lambda>0$
\citep[Chapter 4]{rasmussen2006gpml}. Its weight of order $d$ is
integrable when $d<2\nu$, and the weighted derivative is integrable
for the orders used here, so any
$0<\zeta<\min(1,\nu-r)$ verifies \cref{ass:spectral} when
$\nu>r$. The conditions also hold for positive finite mixtures with
a common smaller exponent. A squared exponential covariance multiplied
by $\cos(\omega_0u)$, with fixed $\omega_0\geq0$, also qualifies
because its spectrum is a positive sum of translated Gaussian densities.

\section{Posterior computation}\label{sec:computation}

Integrating out $\gamma$ gives a posterior density for the remaining
coefficients $\boldsymbol g$. We sample from this density and then draw
$\gamma$ from its conditional truncated normal distribution to recover
the full coefficient vector. We first form the ordinary
Gaussian posterior in \cref{eq:gaussian-posterior} and apply the
appropriate invertible map from
\cref{cor:positive,cor:monotone,cor:convex}, or the common map in
\cref{prop:joint}, to its mean and covariance,
giving the Gaussian laws
\[
 \boldsymbol g\mid\boldsymbol y\sim N(\boldsymbol m_g,\boldsymbol V_g),
 \qquad \gamma\mid\boldsymbol g,\boldsymbol y\sim N(\mu(\boldsymbol g),s^2),
\]
where $\mu$ is affine and $s>0$ is constant in $\boldsymbol g$.
These Gaussian parameters can therefore be prepared before marginal
sampling, together with an invertible factor
$\boldsymbol B\boldsymbol B^T=\boldsymbol V_g$ defining the whitened
coordinates $\boldsymbol g=\boldsymbol m_g+\boldsymbol B\boldsymbol z$,
in which $\boldsymbol z$ has the standard normal law on $\R^m$ under
the ordinary posterior. Mapping through $\boldsymbol B$ preserves the
Gaussian dependence among the remaining coefficients.

At each proposed $\boldsymbol z$, evaluating the marginal density
requires forming $\boldsymbol g$ and its continuum threshold
$\tau_r(\boldsymbol g)$. For the individual cases, this means comparing
both endpoints and all interior stationary candidates from
\cref{cor:positive,cor:monotone,cor:convex}, or obtaining bounds over
the entire interval. The Gaussian log-survival term in
\cref{eq:whitened-target} then weights the standard normal density by
the conditional probability that $\gamma$ satisfies the sign
restriction. Since the observation likelihood is already included
in the ordinary posterior and the overall conditioning probability
is constant in $\boldsymbol z$, neither requires a separate evaluation
during marginal sampling. The monotonicity and
convexity implementations sample this unrestricted $m$-dimensional
density by Hamiltonian Monte Carlo, with the numerical threshold
evaluation described below.

For each retained $\boldsymbol g$, we recover the separated coefficient
by drawing $\gamma$ from
$N(\mu(\boldsymbol g),s^2)$ truncated to
$[\tau_r(\boldsymbol g),\infty)$, using an independent
$U\sim\operatorname{Uniform}(0,1)$
and the inverse-survival formula in \cref{eq:reconstruction}, which
retains its conditional uncertainty. Applying the corresponding inverse
map in \cref{tab:coefficient-reconstruction} then recovers the original
coefficients of $f_{\boldsymbol a}$.
With the exact threshold and draws from the
exact marginal law, these two stages recover the full conditioned
coefficient law, so evaluating the expansion and its derivatives gives
joint posterior curves from which to compute means and pointwise
credible intervals.

For a signed joint restriction satisfying \cref{prop:joint}, the common
map determines both the Gaussian preparation and every quotient
$\eta_r(x;\boldsymbol g)$. At each proposed $\boldsymbol g$, we
evaluate the required continuum and endpoint bounds and take their
maximum to obtain $\tau_A^\varepsilon(\boldsymbol g)$. Since these
bounds all concern the same $\gamma$, a single survival probability
$\Phi^{\mathrm c}((\tau_A^\varepsilon(\boldsymbol g)-\mu(\boldsymbol g))/s)$
weights the marginal density for the entire intersection.
Reconstruction uses this joint lower bound in the truncated normal
and then inverts the common coefficient map. The Gaussian parameters
must be computed under that map, and the individual thresholds in
\cref{tab:cases}, which use different maps, cannot be combined
directly. The supplementary material specifies the HSGP map and
the continuum and endpoint bounds for each joint combination.

In numerical evaluation, a successful threshold calculation returns an
enclosure whose upper endpoint $\widetilde\tau_r$ satisfies
\[
 \tau_r(\boldsymbol g)\leq\widetilde\tau_r(\boldsymbol g)
       \leq\tau_r(\boldsymbol g)+\delta
\]
for the supplied coefficients and a chosen tolerance $\delta>0$.
If marginal weighting and conditional reconstruction use the same
numerical threshold and the enclosure bound holds almost everywhere
under the ordinary posterior, \cref{prop:error} controls the total
variation distance between the resulting target distribution and the
posterior conditioned on the exact sign event.
The conditional standard deviation $s$ provides a natural scale for
choosing $\delta$, because $\delta/s$ bounds the error in the
standardized truncation point. For a prescribed total variation
error $\rho\in(0,1)$, \cref{prop:error} gives the sufficient choice
\[
 \delta\leq\rho\,s\sqrt{2\pi}\,P_{\boldsymbol y}(\mathcal C_r).
\]
In practice, a valid positive lower bound on the conditioning
probability can replace $P_{\boldsymbol y}(\mathcal C_r)$ on the
right-hand side. Without such a bound, a small value of $\delta/s$
controls the standardized threshold error but does not by itself
certify a small posterior error, particularly when the shape event
is rare.

Stable evaluation uses log-survival and inverse-survival calculations
to avoid subtracting nearly equal probabilities in the tails. For
gradient-based sampling, the derivative implementations evaluate
approximate envelope gradients at numerically selected maximizing
locations. At a unique maximum these gradients approximate derivatives of the
exact threshold, while the sampled log values use the rounded upper
threshold. The supplementary material explains how the numerical
values and approximate gradients enter the Stan No-U-Turn sampler
(NUTS), alongside the compact algorithm, Gaussian block formulas,
threshold enclosures, precision controls, and stable reconstruction.

\FloatBarrier
\section{Applications}\label{sec:applications}
We compare ordinary Gaussian and shape-constrained posteriors in two
applications where derivatives have direct scientific interpretations.
Monotonicity ensures nonnegative mercury concentration in a lidar model,
while concavity represents a nonincreasing marginal gain in leaf carbon
assimilation.

\subsection{Atmospheric mercury concentration}\label{sec:lidar}

To illustrate how monotonicity changes derivative inference, we use
all $221$ released mercury lidar observations over
$390$--$720$ meters near the Bella Vista geothermal power station in
Italy, distributed with the R package SemiPar
\citep{holst1996lidar,feng2022sshaped}. Each response is the log ratio
of reflected on-resonant and off-resonant laser signals, with
approximately $100$ pulse pairs averaged during acquisition.
These data have also been analyzed with a globally monotone GP
approximation by \citet{maatouk2025bayesian}.

The shape restriction follows from the differential absorption lidar
(DIAL) approximation, which neglects differential backscatter and
attenuation from other species. Under this approximation, the derivative
of the mean log ratio per meter equals mercury concentration multiplied
by $-1.6\times10^{-5}$ square meters per nanogram. Nonnegative
concentration therefore requires a nondecreasing latent mean for the
negative log ratio, even if the noisy observations are not ordered.

Write $r$ for distance in meters, set $x(r)=(r-390)/330$, and let $Y_i$
be the negative recorded log ratio. We use $128$ basis functions with
center $c=1/2$ and half-width $L=2$, so the computational interval
$[-1.5,2.5]$ contains the physical interval $[0,1]$.
The Mat\'{e}rn $7/2$ spectral prior in \cref{eq:finite-model} gives
the hierarchy
\[
\begin{aligned}
 f_{\boldsymbol a}(x)&=a_0+\sum_{j=1}^{128}a_j\phi_{j,2}(x),
 \qquad a_0\sim N(\mu_0,\sigma_0^2),\\
 a_j&\overset{\mathrm{ind}}\sim
 N\left(0,\sigma_f^2\frac{32\lambda^7}{5(\lambda^2+\omega_j^2)^4}\right),
 \qquad \lambda=\frac{\sqrt7}{\ell},\\
 Y_i\mid\boldsymbol a&\overset{\mathrm{ind}}\sim
 N(f_{\boldsymbol a}(x(r_i)),\sigma_i^2).
\end{aligned}
\]
The intercept is independent of the other coefficients, with prior
mean $\mu_0=0.291156$ and standard deviation $\sigma_0=1.952921$.
These values are the observed response mean and twice its range,
respectively. The underlying
GP has marginal standard deviation $\sigma_f=0.308590$ and normalized
length scale $\ell=0.288869$, corresponding to $95.33$ meters.

Following the noise family of \citet{holst1996lidar}, we allow
measurement variance to change with distance by setting
\[
 \sigma_i=\sigma_{\mathrm{mid}}
 \exp\bigl(\gamma_1(x(r_i)-c)+\gamma_2(x(r_i)-c)^2\bigr).
\]
Here $\sigma_{\mathrm{mid}}=0.056125$ is the noise standard deviation
at the midpoint of $555$ meters, while $\gamma_1=2.330252$ and
$\gamma_2=-0.305387$ describe the linear and quadratic changes in
its logarithm. All standard deviations are on the negative-logratio
scale. We estimate the covariance and noise parameters jointly by
ordinary underlying-GP marginal likelihood and use the same estimates
in both posterior fits.

The inferential targets are the latent negative log ratio
$f_{\boldsymbol{a}}(x(r))$ and the implied concentration
\[
 \frac{1}{330(1.6\times10^{-5})}\left.\frac{\mathrm{d}f_{\boldsymbol{a}}}{\mathrm{d}x}\right|_{x=x(r)}
 \quad\text{nanograms per cubic meter}.
\]
The factor $330$ converts the derivative in normalized coordinates to
a derivative per meter, so conditioning on $\frac{\mathrm{d}f_{\boldsymbol{a}}}{\mathrm{d}x}(x)\geq0$
throughout $[0,1]$ imposes nonnegative concentration over
the observed range. We use the procedure in \cref{sec:computation}
for the monotone posterior and evaluate the unconstrained Gaussian
summaries analytically. Sampler settings, convergence diagnostics, and
the full posterior predictive assessment are given in the supplementary
material.

\cref{fig:lidarcurves} compares concentration inference with and without monotonicity under the same Gaussian model.
At $502.86$ meters, the ordinary posterior mean concentration reaches its minimum on the reconstruction grid at $-42.13$ nanograms per cubic meter, with pointwise $95\%$ equal-tail credible interval $(-94.27,10.00)$ and posterior probability $0.9434$ of a negative concentration.
Conditioning on monotonicity gives a mean concentration of $11.75$ nanograms per cubic meter at that distance, with corresponding interval $(1.10,31.23)$, while changing the response fit little.

\begin{figure}[H]
\centering
\includegraphics[width=\textwidth]{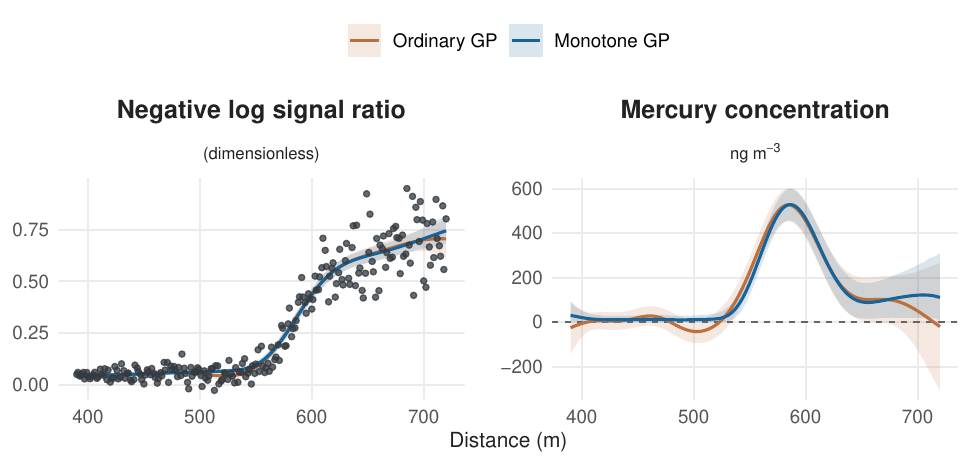}
\caption[Absorption and concentration inference for mercury lidar]{
The left panel shows the latent negative log signal ratio with all $221$ observations as points, and the right shows mercury concentration in nanograms per cubic meter, given under the DIAL approximation by the range derivative of the latent negative log ratio divided by $1.6\times10^{-5}$ square meters per nanogram.
Orange and blue lines denote ordinary and globally monotone posterior means, with shading for pointwise $95\%$ equal-tail latent credible intervals excluding new measurement error.
Both fits use the same rank-$128$ Mat\'{e}rn $7/2$ Gaussian regression model and range-dependent noise variances, with the monotone posterior imposing the restriction throughout $390$--$720$ meters.}
\label{fig:lidarcurves}
\end{figure}

\FloatBarrier
\subsection{Marginal carbon assimilation in a leaf light response}\label{sec:application}

Concavity of a leaf light-response curve means that the marginal gain
in carbon assimilation is nonincreasing with irradiance. We examine
what this assumption contributes to inference from one response sequence
in the Arctic gas-exchange study of \citet{rogers2019arctic}, using
the measurements released by \citet{rogers2019data}. Each record is a
chamber measurement of net assimilation on the same leaf at a specified
incident irradiance. Before fitting either model, we select the smallest
sample identifier in the release, which corresponds to an
Eriophorum angustifolium leaf. The study's quality filters and
exclusion of the dark measurement leave $13$ records over irradiance
$19.1$--$2{,}000.2\,\mu\mathrm{mol}$ quanta $\mathrm m^{-2}\mathrm s^{-1}$.
The measurements were taken in descending illumination over approximately
$37$ minutes, with retained leaf temperatures between $14.1$ and
$15.0\,{}^\circ\mathrm C$.

Let $A(q)$ denote mean net assimilation at irradiance $q$ under the
chamber conditions, measured in $\mu\mathrm{mol}$ $\mathrm{CO}_2$
$\mathrm m^{-2}\mathrm s^{-1}$. We compare the ordinary Gaussian
posterior with its restriction to $\frac{\mathrm{d}^2A}{\mathrm{d}q^2}(q)\leq0$ over the observed
irradiance range. In the scaled coordinate $x=(q-q_0)/d$,
$q_0$ is the midpoint of that range and $d$ its half-width, so the
physical interval becomes $[-1,1]$. The shared model is
\begin{equation}\label{eq:application-model}
\begin{aligned}
 A(q)&=a_0+\sum_{j=1}^{24}a_j\phi_{j,2}((q-q_0)/d),\qquad
 a_0\sim N(0,20^2),\\
 a_j&\overset{\mathrm{ind}}\sim
 N\left(0,\sigma_f^2\sqrt{2\pi}\ell
           \exp(-\ell^2\omega_j^2/2)\right),\\
 y_i\mid A&\overset{\mathrm{ind}}\sim N(A(q_i),\sigma_\varepsilon^2).
\end{aligned}
\end{equation}
The intercept is independent of the other coefficients, and the basis
center is zero in the scaled coordinate. An ordinary Gaussian
marginal-likelihood calculation gives
$\sigma_f=5.625$, $\ell=0.3034$, and $\sigma_\varepsilon=0.3703$,
which both posterior fits use to isolate the effect of the curvature
restriction. The length scale
corresponds to about $301$ irradiance units, and the residual scale has
assimilation units.

The inferential targets are the response $A(q)$, its marginal gain
$\frac{\mathrm{d}A}{\mathrm{d}q}(q)$, and changes in that gain within the observed range.
We report derivatives on the physical irradiance scale, applying factors
$1/d$ and $1/d^2$ to first and second derivatives in the scaled
coordinate. The concavity restriction makes marginal gain
nonincreasing while allowing it to become negative, and imposes no
finite saturation level beyond the fitted interval.

We fit the concave model by applying the convexity sampler to $-A$
and the negated responses, and obtain the unconstrained comparison by
direct Gaussian simulation. Sampling settings, convergence diagnostics,
and posterior predictive assessment are given in the
supplementary material.

\begin{figure}[H]
\centering
\includegraphics[width=\textwidth]{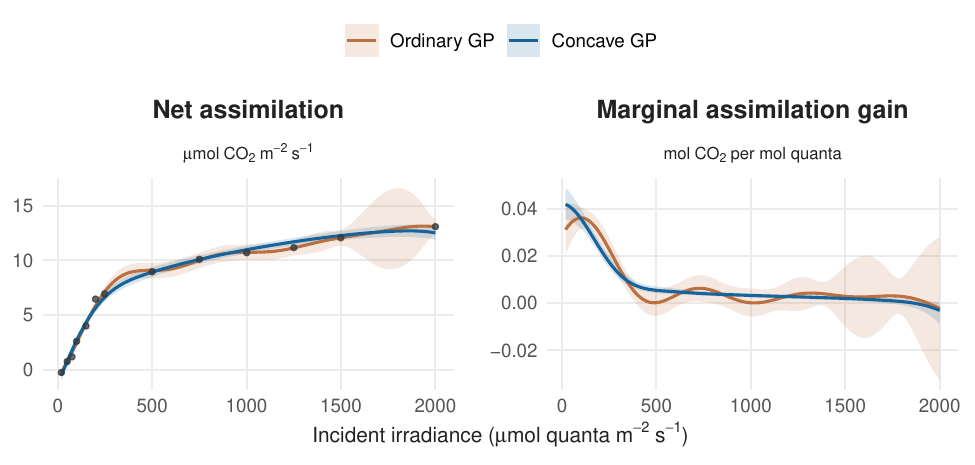}
\caption[Leaf light-response comparison]{Posterior comparison for $13$
net assimilation measurements from one Arctic
Eriophorum angustifolium leaf. The left panel shows net
assimilation in $\mu\mathrm{mol}$ $\mathrm{CO}_2$
$\mathrm m^{-2}\mathrm s^{-1}$, with observations in black.
The right panel shows its derivative with respect to incident irradiance,
in $\mathrm{mol}$ $\mathrm{CO}_2$ per mol quanta.
Orange and blue lines are ordinary Gaussian and concavity-conditioned
posterior means, with shaded pointwise $95\%$ equal-tail credible
intervals for the latent quantities. Both fits use the same covariance
and noise parameters and the same rank-$24$ HSGP expansion.}
\label{fig:light-response}
\end{figure}

The response and derivative panels in \cref{fig:light-response} show
different effects of the constraint. Posterior mean responses are close
at many observed irradiances, but the unconstrained marginal gain can
increase within the range and has substantially wider pointwise intervals
at high irradiance. Concavity imposes nonincreasing marginal gain
on each posterior curve, so the ordering holds for the individual
curves whose uncertainty the right panel summarizes.

\cref{tab:application} gives selected response and derivative summaries.
For example, the assimilation increase from $500$ to $1{,}500$
irradiance units has concave posterior mean \AppGainMean\ and
$95\%$ equal-tail interval $(\AppGainLower,\AppGainUpper)$.
The corresponding unconstrained mean is \OrdGainMean\ with interval
$(\OrdGainLower,\OrdGainUpper)$, all in assimilation units.
The $95\%$ intervals for the difference in marginal gain between $100$
and $1{,}500$ are positive under both fits, so the ordinary Gaussian
analysis already supports diminishing gain for this pair of irradiances.
Conditioning extends that ordering to the entire interval while reducing
posterior uncertainty about marginal gain at high irradiance.
\begin{table}[htbp]
\centering
\small
\begin{tabular}{lrr}
\toprule
Quantity & Unconstrained & Concave \\
\midrule
$A(100)$ & 2.46 (2.11, 2.80) & 2.67 (2.38, 2.97) \\
\addlinespace[2pt]
$A(500)$ & 9.08 (8.38, 9.78) & 8.92 (8.49, 9.38) \\
\addlinespace[2pt]
$A(1{,}500)$ & 12.08 (11.36, 12.80) & 12.27 (11.86, 12.70) \\
\addlinespace[2pt]
$\dfrac{\mathrm{d}A}{\mathrm{d}q}(100)$ & 0.0361 (0.0315, 0.0407) & 0.0357 (0.0316, 0.0399) \\
\addlinespace[2pt]
$\dfrac{\mathrm{d}A}{\mathrm{d}q}(1{,}500)$ & 0.0029 (-0.0099, 0.0155) & 0.0019 (0.0010, 0.0028) \\
\addlinespace[2pt]
$\dfrac{\mathrm{d}A}{\mathrm{d}q}(100)-\dfrac{\mathrm{d}A}{\mathrm{d}q}(1{,}500)$ & 0.0332 (0.0197, 0.0467) & 0.0338 (0.0295, 0.0381) \\
\addlinespace[2pt]
$A(1{,}500)-A(500)$ & 3.00 (2.00, 4.01) & 3.35 (2.73, 3.99) \\
\bottomrule
\end{tabular}

\caption[Response and marginal-gain summaries]{Posterior means with
pointwise $95\%$ equal-tail credible intervals for the leaf light-response
model in \cref{eq:application-model}, with arguments measured in
$\mu\mathrm{mol}$ quanta $\mathrm m^{-2}\mathrm s^{-1}$.
Response values and their differences have units
$\mu\mathrm{mol}$ $\mathrm{CO}_2$ $\mathrm m^{-2}\mathrm s^{-1}$,
while derivatives and their difference have units mol $\mathrm{CO}_2$
per mol quanta. Both columns use the same covariance and noise
estimates, with the concave column additionally conditioning on
nonpositive curvature throughout the observed range.}
\label{tab:application}
\end{table}

The upper endpoint also reveals the limits of a physiological
interpretation. Only \AppEndpointPercent\ of concave draws have a
nonnegative slope at the highest measured irradiance, even though
assimilation increases between the two highest observed irradiances.
The inferred decline near the endpoint depends on the curvature
restriction and stationary prior, and cannot by itself establish
photoinhibition in this leaf.

The example illustrates how a stipulated curvature restriction changes
derivative inference for one sequence, whose observation frame does
not support conclusions about dark respiration or a species response.
Likewise, the analysis does not establish an irradiance optimum
independently of the shape assumption.

\FloatBarrier
\section{Discussion}\label{sec:discussion}

This paper develops a common conditioning construction for positivity,
monotonicity, and convexity in the Laplacian eigenfunction expansion.
The conditioning events themselves are established in constrained GP
regression. Their representation here is specific to the HSGP model:
it uses the existing oscillatory basis and kernel-derived Gaussian
prior to express each whole-interval event through one coefficient,
including joint restrictions when a suitable common direction is
available. This differs from conditioning at selected locations
\citep{daveiga2012inequality,daveiga2020linear} and from choosing a
basis whose coefficient inequalities directly ensure shape
\citep{maatouk2017constraints,lopera2018finite}. The representation
supports joint inference for the curve and its derivatives and
separates the definition of the conditioned posterior from the
numerical evaluation of its threshold.

The approximation theory connects this finite construction to
conditioning the underlying GP. For the individual constraints,
the connection holds under the stated spectral assumptions as both
padding and frequency cutoff increase, giving a basis for using
finite expansions to approximate the corresponding constrained
process. The joint construction is exact within the finite expansion,
while the convergence theorem covers the individual restrictions.

The practical cost of the representation comes from evaluating a
global threshold at each posterior-density calculation. Its
maximizing location can change with the coefficients, creating kinks
that matter for gradient-based sampling even when the marginal law
is strongly log-concave. The threshold-error analysis controls the
posterior change caused by conservative evaluation, while sampler
accuracy and computational cost remain separate considerations.
Developing samplers that exploit this geometry while accommodating
the nonsmooth threshold is a useful direction for further work.

The mercury and leaf applications show that shape restrictions can
substantially change derivative inference even when the fitted response
curves remain similar. These changes matter because the derivatives
determine the inferred mercury concentration and marginal carbon
assimilation, respectively. Comparing the constrained and ordinary
Gaussian posteriors makes the contribution of the shape assumption
explicit in both the estimates and their uncertainty. The construction
thus connects scientifically motivated restrictions to joint inference
for the response and its derivatives, with the chosen constraint
enforced throughout the interval.

\bibliographystyle{plainnat}
\bibliography{references}

\clearpage
\appendix
\def\CombinedProofZone{supplement}
\section{Proofs and supporting results}\label{app:proofs}
\ifdefined\CombinedInlineProofs\else
\subsection{Gaussian conditioning and its specializations}
\printProofs[finite]
\printProofs[positive]
\printProofs[monotone]
\printProofs[convex]
\printProofs[joint]
\subsection{Conservative threshold error}
\printProofs[error]
\subsection{Conditioned posterior approximation}
\printProofs[approximation]
\fi
\subsection{Signed combinations in the Laplacian expansion}\label{app:joint-combinations}
\begin{theoremEnd}[normal,category=combinations]{corollary}[Signed combinations and their reductions]
\CombinedResultCheck{cor:combinations}{corollary}\label{cor:combinations}
Under the finite Gaussian model in \cref{sec:finite-model}, let
$A\subseteq\{0,1,2\}$ be nonempty and fix signs
$\varepsilon_r\in\{-1,1\}$ for $r\in A$.
For $\theta(x)=\omega_1(x-c+L)$, $w(x)=\cos\theta(x)$, and
$J=[w(x_+),w(x_-)]\subset(-1,1)$, define, for a coefficient array
$\boldsymbol b=(b_0,\ldots,b_m)^T$,
\[
 \begin{aligned}
 H_{0,\boldsymbol b}(w)
   &=\sqrt L\,b_0+\sqrt{1-w^2}\sum_{j=1}^m b_jU_{j-1}(w),\\
 H_{1,\boldsymbol b}(w)&=\sum_{j=1}^m jb_jT_j(w),&
 H_{2,\boldsymbol b}(w)&=-\sum_{j=1}^m j^2b_jU_{j-1}(w).
 \end{aligned}
\]
Suppose a fixed $\boldsymbol d=(d_0,\ldots,d_m)^T$ satisfies
$\min_{w\in J}\varepsilon_rH_{r,\boldsymbol d}(w)>0$ for every
$r\in A$. Set $h=d_0+\sum_{j=1}^m d_j\phi_{j,L}$ and choose
$j_*\in\{0,\ldots,m\}$ with $d_{j_*}\ne0$. The coordinates
\[
 \gamma=\frac{a_{j_*}}{d_{j_*}},\qquad
 g_j=a_j-d_j\gamma\quad(j\ne j_*),\qquad g_{j_*}=0
\]
give $\boldsymbol g=(g_j)_{j\ne j_*}\in\R^m$ and
$f_{\boldsymbol g}=g_0+\sum_{j=1}^m g_j\phi_{j,L}$.
Then $f_{\boldsymbol a}=f_{\boldsymbol g}+\gamma h$, with inverse
$a_j=g_j+d_j\gamma$ for $0\leq j\leq m$.
In expressions involving $H_{r,\boldsymbol g}$, the zero coordinate
$g_{j_*}$ is included in the coefficient array.

With $\eta_r$ from \cref{eq:joint-threshold}, define
\[
 \eta_r(x;\boldsymbol g)
   =-\frac{H_{r,\boldsymbol g}(w(x))}{H_{r,\boldsymbol d}(w(x))},
 \qquad M_r(\boldsymbol g)=\max_{x\in I}\eta_r(x;\boldsymbol g),
 \qquad \eta_r^\pm(\boldsymbol g)=\eta_r(x_\pm;\boldsymbol g).
\]
For adjacent selected orders $r,r+1\in A$, let
$E_r(\boldsymbol g)=\eta_r^-(\boldsymbol g)$ when
$\varepsilon_r=\varepsilon_{r+1}$ and
$E_r(\boldsymbol g)=\eta_r^+(\boldsymbol g)$ otherwise.
Suppressing the argument $\boldsymbol g$ on the right, the threshold is
\[
 \tau_A^\varepsilon(\boldsymbol g)=
 \begin{cases}
 M_r, & A=\{r\},\\
 \max\{M_1,E_0\}, & A=\{0,1\},\\
 \max\{M_2,M_0\}, & A=\{0,2\},\ \varepsilon_0=\varepsilon_2,\\
 \max\{M_2,\eta_0^-,\eta_0^+\},
     & A=\{0,2\},\ \varepsilon_0\ne\varepsilon_2,\\
 \max\{M_2,E_1\}, & A=\{1,2\},\\
 \max\{M_2,E_1,E_0\}, & A=\{0,1,2\}.
 \end{cases}
\]
\cref{prop:joint} applies with scalar coefficient
$\gamma=a_{j_*}/d_{j_*}$. Thus
$f_{\boldsymbol a}\in\mathcal C_A^\varepsilon$ exactly when
$\gamma\geq\tau_A^\varepsilon(\boldsymbol g)$.

Each maximum $M_r$ is attained at a physical endpoint or a location
whose cosine coordinate is an interior root in $J$ of
\[
 H_{r,\boldsymbol d}(w)\frac{\mathrm{d}H_{r,\boldsymbol g}}{\mathrm{d}w}(w)
 -H_{r,\boldsymbol g}(w)\frac{\mathrm{d}H_{r,\boldsymbol d}}{\mathrm{d}w}(w)=0.
\]
For $r=1,2$, the left side is a polynomial.
All root multiplicities and reduced degrees are included.
If the left side vanishes identically, the quotient is constant and
either endpoint suffices.

These formulas cover the 26 nonempty combinations without opposing
signs, with positive ordinary posterior probability whenever the
stated direction condition holds for the selected combination.
The remaining 37 selections contain opposing signs. At orders zero,
one, and two these force, respectively, a zero, constant, or affine
function, subject to any further restrictions. In the finite
expansion, affine curves are constant, so all 37 events have
ordinary posterior probability zero.
\end{theoremEnd}

\begin{proofEnd}
\CombinedProofCheck{cor:combinations}
Fix the selected orders, signs, and direction coefficients satisfying
\cref{cor:combinations}. The inverse map follows by substitution, so
the coordinates preserve the full coefficient space. The Chebyshev identities in
\cref{def:chebyshev} give
\[
 \begin{aligned}
 f_{\boldsymbol b}(x)&=\frac{H_{0,\boldsymbol b}(w(x))}{\sqrt L},\\
 \frac{\mathrm{d}f_{\boldsymbol b}}{\mathrm{d}x}(x)
    &=\frac{\omega_1}{\sqrt L}H_{1,\boldsymbol b}(w(x)),\\
 \frac{\mathrm{d}^2f_{\boldsymbol b}}{\mathrm{d}x^2}(x)
    &=\frac{\omega_1^2\sin\theta(x)}{\sqrt L}
       H_{2,\boldsymbol b}(w(x)).
 \end{aligned}
\]
Since $\sin\theta$ is positive on compact $I$, the assumed conditions
on $H_{r,\boldsymbol d}$ give the uniform signed derivative bounds
required by \cref{prop:joint}. The positive factors cancel from the
ratios, yielding the displayed expression for $\eta_r$ and the
conditioned posterior.

A signed slope restriction makes the minimum of
$\varepsilon_0 f_{\boldsymbol a}$ occur at $x_-$ when
$\varepsilon_0=\varepsilon_1$, and at $x_+$ otherwise.
This gives $E_0$. Applying the same argument to the signed slope
$\varepsilon_1\frac{\mathrm{d}f_{\boldsymbol a}}{\mathrm{d}x}$
under the curvature restriction gives $E_1$. When
$\varepsilon_0\ne\varepsilon_2$, the signed curve
$\varepsilon_0 f_{\boldsymbol a}$ is concave, so its nonnegativity
is equivalent to nonnegative endpoint values by the chord inequality.
When the signs agree, the signed curve is convex and its minimum may
be interior, so the value restriction retains $M_0$. Together with
the continuum slope and curvature bounds $M_1$ and $M_2$, these
equivalences give every displayed threshold, because each pointwise
restriction is $\gamma\geq\eta_r(x;\boldsymbol g)$.

The cosine coordinate is strictly decreasing on $I$, and the signed
denominators stay positive on compact $J$. The extreme value theorem,
Fermat's theorem, and the quotient rule therefore give the stated
endpoint and stationary candidates. The functions $H_{r,\boldsymbol b}$
are analytic on $(-1,1)$, so a nonzero stationary expression has only
finitely many roots on $J$. If it vanishes identically, the mean value
theorem gives constancy of the quotient.

Choosing no restriction or one of two signs at each order gives
$3^3-1=26$ nonempty combinations: six singles, twelve pairs, and eight
triples. Of the $2^6-1=63$ nonempty selections from the six constraints,
the other 37 contain opposing signs, which force the corresponding
derivative to vanish identically. This gives zero, a constant, or an
affine function. Analyticity extends any affine identity on $I$ to
the padded interval, whose two endpoint values both equal $a_0$, so
the affine curve is constant. Linear independence of the sine basis
gives $a_1=\cdots=a_m=0$, with $a_0=0$ as well when both value signs
are imposed. These proper linear subspaces have zero probability
under the nondegenerate Gaussian posterior.
\end{proofEnd}

\begingroup
\small
\setlength{\tabcolsep}{3pt}
\setlength{\LTcapwidth}{\textwidth}
\begin{longtable}{@{}llll@{}}
\caption{Implementing the 20 joint combinations without opposing
constraints under \cref{cor:combinations}.
P and N denote nonnegativity and nonpositivity, I and D denote
nondecreasing and nonincreasing curves, and CX and CV denote convexity
and concavity. Choose fixed direction coefficients $\boldsymbol d$
so that every signed profile in the second column is strictly positive
throughout $J$. For a pivot $d_{j_*}\ne0$, use
$\gamma=a_{j_*}/d_{j_*}$ and $g_j=a_j-d_j\gamma$, so that
$g_{j_*}=0$ and $a_j=g_j+d_j\gamma$ for $0\leq j\leq m$.
Every displayed coefficient bound in a row must hold, giving
$\gamma\geq\tau_A^\varepsilon(\boldsymbol g)$ by taking the largest
displayed right-hand side. With arguments $\boldsymbol g$ suppressed,
$M_r=\max_{x\in I}\eta_r(x;\boldsymbol g)$ requires an interval maximum,
evaluated at the endpoint and stationary candidates in
\cref{cor:combinations}, whereas
$\eta_r^\pm=\eta_r(x_\pm;\boldsymbol g)$ requires only evaluation
at the left endpoint $x_-$ or right endpoint $x_+$.
All ratios
$\eta_r(x;\boldsymbol g)=-H_{r,\boldsymbol g}(w(x))/H_{r,\boldsymbol d}(w(x))$
use the same direction and coefficient map within a row.
A dash indicates no separate endpoint bound.}
\label{tab:constraint-combinations}\\
\toprule
Combination & Direction condition on $J$ & Continuum bound & Endpoint bound\\
\midrule
\endfirsthead
\multicolumn{4}{@{}l}{\tablename\ \thetable\ (continued)}\\
\toprule
Combination & Direction condition on $J$ & Continuum bound & Endpoint bound\\
\midrule
\endhead
\midrule
\multicolumn{4}{r@{}}{Continued on the next page}\\
\endfoot
\bottomrule
\endlastfoot
\multicolumn{4}{@{}l}{\textit{Pairs of constraints}}\\*
P, I & $H_{0,\boldsymbol d},H_{1,\boldsymbol d} >0$ &
  $\gamma\geq M_1$ & $\gamma\geq\eta_0^-$\\
\addlinespace[3pt]
P, D & $H_{0,\boldsymbol d},-H_{1,\boldsymbol d} >0$ &
  $\gamma\geq M_1$ & $\gamma\geq\eta_0^+$\\
\addlinespace[3pt]
N, I & $-H_{0,\boldsymbol d},H_{1,\boldsymbol d} >0$ &
  $\gamma\geq M_1$ & $\gamma\geq\eta_0^+$\\
\addlinespace[3pt]
N, D & $-H_{0,\boldsymbol d},-H_{1,\boldsymbol d} >0$ &
  $\gamma\geq M_1$ & $\gamma\geq\eta_0^-$\\
\addlinespace[3pt]
P, CX & $H_{0,\boldsymbol d},H_{2,\boldsymbol d} >0$ &
  $\gamma\geq\max\{M_2,M_0\}$ & ---\\
\addlinespace[3pt]
P, CV & $H_{0,\boldsymbol d},-H_{2,\boldsymbol d} >0$ &
  $\gamma\geq M_2$ & $\gamma\geq\max\{\eta_0^-,\eta_0^+\}$\\
\addlinespace[3pt]
N, CX & $-H_{0,\boldsymbol d},H_{2,\boldsymbol d} >0$ &
  $\gamma\geq M_2$ & $\gamma\geq\max\{\eta_0^-,\eta_0^+\}$\\
\addlinespace[3pt]
N, CV & $-H_{0,\boldsymbol d},-H_{2,\boldsymbol d} >0$ &
  $\gamma\geq\max\{M_2,M_0\}$ & ---\\
\addlinespace[3pt]
I, CX & $H_{1,\boldsymbol d},H_{2,\boldsymbol d} >0$ &
  $\gamma\geq M_2$ & $\gamma\geq\eta_1^-$\\
\addlinespace[3pt]
I, CV & $H_{1,\boldsymbol d},-H_{2,\boldsymbol d} >0$ &
  $\gamma\geq M_2$ & $\gamma\geq\eta_1^+$\\
\addlinespace[3pt]
D, CX & $-H_{1,\boldsymbol d},H_{2,\boldsymbol d} >0$ &
  $\gamma\geq M_2$ & $\gamma\geq\eta_1^+$\\
\addlinespace[3pt]
D, CV & $-H_{1,\boldsymbol d},-H_{2,\boldsymbol d} >0$ &
  $\gamma\geq M_2$ & $\gamma\geq\eta_1^-$\\
\addlinespace
\multicolumn{4}{@{}l}{\textit{Triples of constraints}}\\*
P, I, CX & $H_{0,\boldsymbol d},H_{1,\boldsymbol d},H_{2,\boldsymbol d} >0$ &
  $\gamma\geq M_2$ & $\gamma\geq\max\{\eta_1^-,\eta_0^-\}$\\
\addlinespace[3pt]
P, I, CV & $H_{0,\boldsymbol d},H_{1,\boldsymbol d},-H_{2,\boldsymbol d} >0$ &
  $\gamma\geq M_2$ & $\gamma\geq\max\{\eta_1^+,\eta_0^-\}$\\
\addlinespace[3pt]
P, D, CX & $H_{0,\boldsymbol d},-H_{1,\boldsymbol d},H_{2,\boldsymbol d} >0$ &
  $\gamma\geq M_2$ & $\gamma\geq\max\{\eta_1^+,\eta_0^+\}$\\
\addlinespace[3pt]
P, D, CV & $H_{0,\boldsymbol d},-H_{1,\boldsymbol d},-H_{2,\boldsymbol d} >0$ &
  $\gamma\geq M_2$ & $\gamma\geq\max\{\eta_1^-,\eta_0^+\}$\\
\addlinespace[3pt]
N, I, CX & $-H_{0,\boldsymbol d},H_{1,\boldsymbol d},H_{2,\boldsymbol d} >0$ &
  $\gamma\geq M_2$ & $\gamma\geq\max\{\eta_1^-,\eta_0^+\}$\\
\addlinespace[3pt]
N, I, CV & $-H_{0,\boldsymbol d},H_{1,\boldsymbol d},-H_{2,\boldsymbol d} >0$ &
  $\gamma\geq M_2$ & $\gamma\geq\max\{\eta_1^+,\eta_0^+\}$\\
\addlinespace[3pt]
N, D, CX & $-H_{0,\boldsymbol d},-H_{1,\boldsymbol d},H_{2,\boldsymbol d} >0$ &
  $\gamma\geq M_2$ & $\gamma\geq\max\{\eta_1^+,\eta_0^-\}$\\
\addlinespace[3pt]
N, D, CV & $-H_{0,\boldsymbol d},-H_{1,\boldsymbol d},-H_{2,\boldsymbol d} >0$ &
  $\gamma\geq M_2$ & $\gamma\geq\max\{\eta_1^-,\eta_0^-\}$\\
\end{longtable}
\endgroup

\subsection{Whitened posterior geometry}
\begin{theoremEnd}[normal,category=geometry]{proposition}[Strong log-concavity of the marginal posterior]
\CombinedResultCheck{prop:geometry}{proposition}\label{prop:geometry}
Under \cref{thm:finite}, let
$\boldsymbol B\boldsymbol B^T=\boldsymbol V_g$ with $\boldsymbol B$
invertible and put $\boldsymbol g=\boldsymbol m_g+\boldsymbol B\boldsymbol z$.
The induced density $q_r$ of $\boldsymbol z$ is $1$-strongly
log-concave. That is, $\log q_r(\boldsymbol z)+\|\boldsymbol z\|^2/2$
is concave. It has a unique mode and finite moments of every
polynomial order.
\end{theoremEnd}
\begin{proofEnd}
\CombinedProofCheck{prop:geometry}
Fix the Gaussian posterior, selected order, and invertible whitening
factor in \cref{prop:geometry}. Since $\tau_r$ is a maximum of
linear functions, it is convex. Subtracting the affine conditional
mean and dividing by $s>0$ therefore makes
$\alpha_r=(\tau_r-\mu)/s$ convex in the whitened coordinates.
Under the ordinary posterior,
$\eta=(\gamma-\mu(\boldsymbol g))/s$ is standard normal and independent
of $\boldsymbol z$. The joint standard normal density in
$(\boldsymbol z,\eta)$ restricted to the convex epigraph
$\eta\geq\alpha_r(\boldsymbol z)$ is $1$-strongly log-concave and
has positive normalizer by \cref{thm:finite}.
Preservation under marginalization
\citep[Theorem 3.8]{saumard2014logconcavity} gives the same constant
for $q_r$.
To obtain the moment and mode conclusions, use the survival factor's
continuity, positivity, and upper bound of one:
$q_r$ is continuous and bounded above by the standard normal density
divided by the positive event probability. This bound gives all
polynomial moments and convergence of the density to zero at infinity,
so continuity implies attainment of a mode, which strong
log-concavity makes unique.
\end{proofEnd}

\FloatBarrier
\section{Numerical evaluation}\label{app:numerical-evaluation}\label{sec:numerics}

The posterior construction requires three numerical operations. We
first prepare the ordinary Gaussian posterior, then evaluate the
shape threshold whenever the sampler proposes a coefficient vector,
and finally draw the omitted coefficient conditionally on each
retained vector. The Gaussian preparation is performed once, whereas
threshold evaluation is repeated throughout sampling. The main
numerical issue is therefore to bound a maximum over the entire
constraint interval reliably and to use the same bound in the
marginal density and conditional reconstruction.

\subsection{Gaussian preparation and the common algorithm}
\label{subsec:gaussian-preparation}

The preparation separates the coefficients that will be sampled
from the scalar that will be integrated out. Let $\boldsymbol A$ be
the invertible linear map from the original coefficients to
$(\boldsymbol g,\gamma)$ under the applicable conditions in
\cref{cor:positive,cor:monotone,cor:convex}, or the common map in
\cref{cor:combinations} for a joint restriction. The ordinary posterior
then has mean $\boldsymbol A\boldsymbol m$ and covariance
$\boldsymbol A\boldsymbol V\boldsymbol A^T$. Denote the mean blocks
by $\boldsymbol m_g$ and $m_\gamma$, the marginal covariance of
$\boldsymbol g$ by $\boldsymbol V_g$, the variance of $\gamma$ by
$v_\gamma$, and their cross-covariance by
$\boldsymbol v=\Cov(\boldsymbol g,\gamma\mid\boldsymbol y)$.
Gaussian conditioning gives
\begin{equation}\label{eq:gaussian-blocks}
 \mu(\boldsymbol g)=m_\gamma+
      \boldsymbol v^T\boldsymbol V_g^{-1}(\boldsymbol g-\boldsymbol m_g),
 \qquad s^2=v_\gamma-\boldsymbol v^T\boldsymbol V_g^{-1}\boldsymbol v.
\end{equation}
These formulas give the conditional mean and variance of $\gamma$
for any value of $\boldsymbol g$. To put the sampled coefficients
on a common Gaussian scale, choose an invertible factor
$\boldsymbol B\boldsymbol B^T=\boldsymbol V_g$ and set
$\boldsymbol g(\boldsymbol z)=\boldsymbol m_g+\boldsymbol B\boldsymbol z$.
Then \cref{eq:gaussian-blocks} gives
$\mu(\boldsymbol g(\boldsymbol z))=m_\gamma+\boldsymbol\lambda^T\boldsymbol z$,
where $\boldsymbol\lambda=\boldsymbol B^{-1}\boldsymbol v$ is
computed by a linear solve. Under the ordinary posterior,
$\boldsymbol z$ has a standard normal distribution. The shape
restriction then enters through the survival weight in
\cref{eq:whitened-target}.

When computing $s^2$, subtracting two nearly equal terms in
\cref{eq:gaussian-blocks} can lose precision. The equivalent formula
$s^2=(\boldsymbol d^T\boldsymbol V^{-1}\boldsymbol d)^{-1}$,
where $\boldsymbol d$ is the column of $\boldsymbol A^{-1}$
corresponding to $\gamma$, avoids this subtraction.
The quadratic form is the $\gamma$ diagonal entry of the transformed
precision matrix and can be evaluated using the factored posterior
precision without forming $\boldsymbol V^{-1}$ explicitly.

The ordinary Gaussian posterior can itself be prepared using matrix
factorizations and linear solves. For a general noise covariance
$\boldsymbol\Sigma$, solves with its Cholesky factor transform the
observation errors to have identity covariance. We apply the same transformation
to the design and to the response centered at
$\boldsymbol X\boldsymbol m_0$, then scale the design columns by
the prior standard deviations. In these coordinates, the prior
precision is the identity. Stacking the transformed design above an
identity matrix and applying a QR factorization therefore gives a
triangular factor of the posterior precision. Linear solves supply
the posterior mean and covariance factors, after which the full
transformation $\boldsymbol A\boldsymbol V\boldsymbol A^T$
preserves the dependence introduced by the coefficient map.

\cref{alg:posterior} separates the repeated marginal-density
evaluation from reconstruction after sampling. Both stages use the
same threshold rule, tolerance, and precision policy, so they describe
the same conservative conditioning event. A gradient-based sampler
also needs a vector to guide its proposals, called the force below.
We obtain it from the envelope derivative in
\cref{subsec:threshold-derivatives}, which differentiates the
maximized objective without differentiating the numerical search
for its maximizing location.

\begin{algorithm}[htbp]
\caption{Sign-constrained marginal sampling and reconstruction.
For a joint event satisfying \cref{prop:joint}, use its common
coefficient map and replace $\tau_r$ by $\tau_A^\varepsilon$
throughout, including the upper threshold used in both stages.}
\label{alg:posterior}
\begin{algorithmic}[1]
\Require Gaussian model, event $\mathcal C_r$ or $\mathcal C_A^\varepsilon$, admissible map, interval $I$
\Require Absolute tolerance $\delta>0$ and precision budget
\State Form the ordinary posterior and apply the invertible shape-specific map
\State Prepare $\boldsymbol m_g,\boldsymbol B,m_\gamma,\boldsymbol\lambda,s$
\Function{Marginal evaluation}{$\boldsymbol z$}
 \State Set $\boldsymbol g=\boldsymbol m_g+\boldsymbol B\boldsymbol z$
 \State Set $\mu=m_\gamma+\boldsymbol\lambda^T\boldsymbol z$
 \State Bound $\tau_r(\boldsymbol g)$ between $\ell$ and $u$ with $u-\ell\leq\delta$, or report failure
 \State Set $\widetilde\tau_r=u$ and $\widetilde\alpha=(\widetilde\tau_r-\mu)/s$
 \State Return $-\|\boldsymbol z\|^2/2+\log\Phi^{\mathrm c}(\widetilde\alpha)$ and the envelope force
\EndFunction
\State Use the marginal sampler to retain draws of $\boldsymbol z$
\For{each retained $\boldsymbol z$}
 \State Set $\boldsymbol g=\boldsymbol m_g+\boldsymbol B\boldsymbol z$ and $\mu=m_\gamma+\boldsymbol\lambda^T\boldsymbol z$
 \State Evaluate the same upper-threshold rule and set $\widetilde\alpha=(\widetilde\tau_r-\mu)/s$
 \State Draw $U\sim\operatorname{Uniform}(0,1)$ independently
 \State Compute $\eta=-\Phi^{-1}(U\Phi^{\mathrm c}(\widetilde\alpha))$ using log probabilities
 \State Set $\gamma=\widetilde\tau_r+s(\eta-\widetilde\alpha)$ and invert the coefficient map
 \State Evaluate the reconstructed curve and requested functionals
\EndFor
\end{algorithmic}
\end{algorithm}

\subsection{Conservative threshold evaluation}
\label{subsec:threshold-computation}

The threshold is a maximum over the entire constraint interval.
Evaluating its objective at finitely many locations gives a lower
bound, while an upper bound must also control values between those
locations. The calculation therefore maintains both bounds and
refines them until their difference is within the requested tolerance.

\paragraph{Bounds used for conditioning.}
The routine takes the remaining coefficients in their declared order,
the shape case and basis geometry, a positive absolute tolerance
$\delta$ in coefficient units, and an arithmetic precision limit.
A successful evaluation returns
\[
 \ell\leq\tau_r(\boldsymbol g)\leq u,
 \qquad u-\ell\leq\delta,
\]
together with an approximate maximizing location for the envelope
force. Choosing $\widetilde\tau_r=u$ then ensures
$0\leq\widetilde\tau_r-\tau_r\leq\delta$. The resulting lower bound on
$\gamma$ can be slightly too large, but cannot admit a curve that
violates the constraint. If the requested enclosure cannot be
obtained, the routine reports failure.

\paragraph{Locating and evaluating extrema.}
For each individual constraint, the maximum occurs at an endpoint
or an interior stationary location.
\cref{cor:positive,cor:monotone,cor:convex} express the interior
candidates as roots of a polynomial in the cosine coordinate $w\in J$.
Real-root isolation finds disjoint intervals containing every
distinct real root and establishes that none have been omitted
\citep{johansson2017arb}. The represented coefficients are treated as
exact rationals, allowing repeated factors to be removed before
isolation while retaining roots of even multiplicity. This arithmetic
also detects reduced degree without discarding small nonzero
coefficients. A constant stationary polynomial requires only endpoint
evaluations, including when it is identically zero.

Interval arithmetic then encloses the objective at the endpoints
and over each root interval. Rounding is directed outward so that
these enclosures include arithmetic error. The largest candidate
upper bound gives $u$, while valid lower bounds come from endpoints
and roots known to belong to $J$. Refining the candidate enclosures
reduces the gap between these bounds until it is at most $\delta$.

Roots near the boundary of $J$ need additional care. A root interval
known to lie outside $J$ is discarded. One crossing the boundary
remains an upper-bound candidate, but supplies a lower bound only
after its root is shown to belong to $J$. The calculation can still
stop without resolving that location if the overall gap is already
within $\delta$. For quotient objectives, division also requires a
strictly positive interval lower bound for the denominator.

\paragraph{Evaluation for the individual constraints.}
For positivity, isolate the roots of $P_{0,\boldsymbol g}$ and
evaluate the objective $-f_{\boldsymbol g}$ over the corresponding
physical locations. Interval evaluation of this trigonometric
expression encloses the candidate values.

For monotonicity, the objective is $-P_{\boldsymbol g}/D$, with
stationary polynomial
\[
 \frac{\mathrm{d}P_{\boldsymbol g}}{\mathrm{d}w}D
 -P_{\boldsymbol g}\frac{\mathrm{d}D}{\mathrm{d}w}.
\]
The polynomials in \cref{eq:monotone-polynomials} use every spectral
coefficient, with $g_2=0$ and the intercept excluded, and satisfy
$D\geq1$ on $J$. The lidar implementation first bounds the equivalent
trigonometric expression, retaining all harmonics, and uses root
isolation as a fallback.

For convexity, the polynomial objective $Q_{\boldsymbol g}$ can be
bounded before finding its stationary points. Map $J$ affinely to
$[0,1]$ and express the polynomial in the Bernstein basis. These
basis functions are nonnegative and sum to one, so every polynomial
value is a weighted average of the coefficients and lies between
their smallest and largest values. Ball arithmetic, which represents
each quantity by a midpoint and an error radius, encloses the change
of basis and the cosine endpoints.

De Casteljau subdivision rewrites the polynomial on smaller
intervals, with outward rounding preserving the bounds. Repeatedly
subdividing the interval with the largest upper bound refines $u$,
while endpoint and midpoint evaluations improve $\ell$.
If the gap still exceeds $\delta$ when the subdivision budget is
reached, isolate the roots of the exact rational derivative of
$Q_{\boldsymbol g}$ and increase arithmetic precision as needed.

\paragraph{Joint restrictions.}
For a signed joint event, use a common coefficient map satisfying
\cref{prop:joint}. Every selected order then shares the same direction
$h$ and remaining coefficients $\boldsymbol g$, so its quotient
objective must be formed from \cref{eq:joint-threshold} under this map.
Subdivide $I$ and bound each quotient on every subinterval, verifying
a positive lower bound for its signed denominator before division.
Evaluations at points in $I$ supply lower bounds.

If $[\ell_r,u_r]$ encloses the maximum for order $r\in A$, then
\[
 \max_{r\in A}\ell_r
 \leq\tau_A^\varepsilon(\boldsymbol g)
 \leq\max_{r\in A}u_r.
\]
The width of this joint enclosure is at most
$\max_{r\in A}(u_r-\ell_r)$, so requiring each width to be at most
$\delta$ suffices. The reductions in
\cref{cor:combinations,tab:constraint-combinations} replace some
continuum searches by endpoint evaluations. Enclose each retained
maximum and endpoint quotient with outward rounding, then take the
maxima of their lower and upper bounds. The same width guarantee
applies. The positive--convex and negative--concave pairs retain the
continuum value threshold $M_0$ alongside the curvature threshold
$M_2$.

\paragraph{Precision and error control.}
These guarantees concern the numerical coefficients and geometry
supplied to the routine, treated as exact inputs. They exclude errors
from the preceding Gaussian preparation. Applying \cref{prop:error}
requires the threshold-error bound to hold almost everywhere under
the ordinary Gaussian law.

Fixing the precision policy and the rules for choosing among
candidate locations makes repeated evaluation deterministic.
Nonfinite inputs and enclosures unresolved at the precision limit
are reported as failures. The Gaussian survival weight is positive
at every finite coefficient vector, so assigning zero density to an
unresolved evaluation would impose an additional restriction.
Such failures therefore cannot be silently treated as rejected
coefficient values.

\subsection{Threshold derivatives}
\label{subsec:threshold-derivatives}

Gradient-based sampling requires the derivative of the log marginal
density with respect to the sampled coefficients. The shape restriction
enters through the Gaussian survival weight in
\cref{eq:whitened-target}, so this calculation depends on how the
threshold changes with the coefficients. The maximum representation
of the threshold gives these derivatives from its maximizing locations,
without differentiating the numerical search used to find them.

\paragraph{Slopes at maximizing locations.}
At a fixed $x$, the threshold objective is linear in
$\boldsymbol g$. Under \cref{thm:finite}, write
\[
 \tau_r(\boldsymbol g)
 =\max_{x\in I}\boldsymbol\beta_r(x)^T\boldsymbol g,
\]
where $\boldsymbol\beta_r(x)$ contains the coefficient multipliers,
ordered as the entries of $\boldsymbol g$. The positive denominator
and compact interval make this vector continuous and bounded in $x$.
Consequently, $\tau_r$ is convex, positively homogeneous, and
Lipschitz, with Lipschitz constant at most
$\max_{x\in I}\|\boldsymbol\beta_r(x)\|$.

Let $\mathcal A_r(\boldsymbol g)$ denote the locations attaining
the maximum, which we call active locations.
For a perturbation in a direction $\boldsymbol b\in\R^m$,
Danskin's theorem \citep[Exercise 6.7]{bertsekas2010convexchapter}
gives
\[
 \left.\frac{\mathrm{d}}{\mathrm{d}u}
       \tau_r(\boldsymbol g+u\boldsymbol b)\right|_{u=0+}
 =\max_{x\in\mathcal A_r(\boldsymbol g)}
       \boldsymbol\beta_r(x)^T\boldsymbol b.
\]
The theorem applies because $I$ is compact and the coefficient
gradients are continuous. Its formula compares the slopes at the
currently maximizing locations and takes the largest slope in the
chosen direction.

The subdifferential consists of all convex combinations of the active
vectors $\boldsymbol\beta_r(x)$. A gradient exists exactly when all
these vectors agree, in which case their common value is the column
gradient. Thus a unique maximizing location suffices for
differentiability. Several locations also permit a gradient if they
give the same vector, while different active vectors produce a kink.

\paragraph{Individual and joint constraints.}
For the individual constraints, the coefficient vectors follow
directly from the objectives in
\cref{cor:positive,cor:monotone,cor:convex}. With
$w(x)=\cos(\omega_1(x-c+L))$, their spectral components are
\[
 \begin{aligned}
  \beta_{0,j}(x)&=-\phi_{j,L}(x),
       &&1\leq j\leq m,\\
  \beta_{1,j}(x)&=-jT_j(w(x))/D(w(x)),
       &&1\leq j\leq m,\quad j\ne2,\\
  \beta_{2,j}(x)&=j^2U_{j-1}(w(x)),
       &&2\leq j\leq m.
 \end{aligned}
\]
The intercept components in the monotonicity and convexity coordinates
are zero. Where the threshold is differentiable, these formulas
evaluate its gradient at any maximizing location.

For the joint threshold in \cref{prop:joint}, form the vectors from
the common coefficient map so that
$\eta_r(x;\boldsymbol g)=\boldsymbol\beta_r(x)^T\boldsymbol g$.
An active candidate is now a pair $(r,x)$ attaining the joint
threshold. The directional derivative is the largest
$\boldsymbol\beta_r(x)^T\boldsymbol b$ among these pairs, and a
gradient exists exactly when their vectors agree. This includes ties
between different derivative orders. The reductions in
\cref{cor:combinations} preserve the threshold, so the same calculation
can use the retained continuum objectives and endpoint quotients.

\paragraph{Gradient of the marginal log density.}
Let $q_r(\boldsymbol z)$ be the exact conditioned density in whitened
coordinates from \cref{eq:whitened-target}. Define the standardized
threshold by
\[
 \alpha_r(\boldsymbol z)
 =\frac{\tau_r(\boldsymbol g(\boldsymbol z))
          -\mu(\boldsymbol g(\boldsymbol z))}{s},
\]
and let $\varphi_1$ denote the standard normal density.
At a point where the active coefficient vectors agree, choose any
maximizing location $x_\star$. The map
$\boldsymbol g(\boldsymbol z)=\boldsymbol m_g+\boldsymbol B\boldsymbol z$
transforms the threshold gradient to
$\boldsymbol B^T\boldsymbol\beta_r(x_\star)$, while the conditional
mean has gradient $\boldsymbol\lambda$. The chain rule therefore gives
the column gradients
\begin{equation}\label{eq:target-gradient}
 \begin{aligned}
 \frac{\partial\alpha_r}{\partial\boldsymbol z}
   &=\frac{\boldsymbol B^T\boldsymbol\beta_r(x_\star)-\boldsymbol\lambda}{s},\\
 \frac{\partial\log q_r}{\partial\boldsymbol z}
   &=-\boldsymbol z-
       \frac{\varphi_1(\alpha_r)}{\Phi^{\mathrm c}(\alpha_r)}
       \frac{\partial\alpha_r}{\partial\boldsymbol z}.
 \end{aligned}
\end{equation}
The term $-\boldsymbol z$ comes from the standard normal density,
while the second term differentiates the log survival probability.
Its weight $\varphi_1(\alpha_r)/\Phi^{\mathrm c}(\alpha_r)$ is the
inverse Mills ratio. The gradient of $\alpha_r$ accounts for changes
in both the threshold and the conditional mean of $\gamma$.

\paragraph{The numerical force.}
The numerical adapter selects an approximate maximizing location
deterministically, evaluates $\boldsymbol\beta_r$ there, and uses
$\widetilde\alpha=(\widetilde\tau_r-\mu)/s$ in the normal-tail factor.
Substituting these quantities in the envelope formula gives the force
used by the sampler.

The threshold tolerance controls objective values, but locations with
almost equal values can have substantially different coefficient
slopes. A narrow enclosure of the maximum therefore does not guarantee
an accurate force near a change of maximizing location. The numerical
log density uses the upper threshold, so the returned envelope force need
not be its exact gradient. \cref{app:stan-implementation} explains how
this force is used in sampling.

\subsection{Stable tail evaluation and reconstruction}
\label{subsec:tail-computation}

When the threshold lies far above the scalar conditional mean, its
Gaussian survival probability can be extremely small. Computing
$1-\Phi(\alpha)$ can then lose precision or return zero, even
though the probability is positive. We instead evaluate
$\log\Phi^{\mathrm c}(\alpha)$ directly as $\log\Phi(-\alpha)$ using a
normal log-tail routine. The inverse Mills ratio
in \cref{eq:target-gradient} is then
$\exp(\log\varphi_1(\alpha)-\log\Phi^{\mathrm c}(\alpha))$.

For reconstruction, draw $U\sim\operatorname{Uniform}(0,1)$
independently and use a normal quantile routine that accepts log
probabilities to solve
\[
 \log\Phi(-\eta)=\log U+\log\Phi^{\mathrm c}(\widetilde\alpha).
\]
In exact arithmetic this gives $\eta>\widetilde\alpha$ for
$0<U<1$, and
$\gamma=\widetilde\tau_r+s(\eta-\widetilde\alpha)$ equals the
truncated-normal draw in \cref{eq:reconstruction} with the upper
threshold substituted. Expressing the draw relative to the threshold
is useful when the conditional mean and a large positive normal
quantile would nearly cancel. A nonfinite quantile or a
computed $\eta<\widetilde\alpha$ is a reconstruction failure.
These arithmetic safeguards do not extend \cref{prop:error} to
rounding in Gaussian preparation or tail evaluation, since that
bound assumes exact Gaussian and tail operations.

\FloatBarrier
\section{Implementation in Stan}\label{app:stan-implementation}

For monotonicity and convexity, the Stan programs implement
\cref{alg:posterior} using the No-U-Turn sampler (NUTS)
\citep{hoffman2014nuts,carpenter2017stan}. The sampled parameter is
the unconstrained vector $\boldsymbol z\in\mathbb R^m$. Gaussian
preparation supplies $\boldsymbol m_g,\boldsymbol B,m_\gamma,
\boldsymbol\lambda$, and $s>0$ as data, together with the basis
geometry, constraint interval, and threshold tolerance. The threshold
routine also has a specified precision budget. The rank and padding satisfy the conditions of
\cref{cor:monotone,cor:convex}. Preparation allows a general
positive-definite noise covariance. The convexity program uses a basis
centered at zero, so a model centered at $c$ is represented by
subtracting $c$ from both observation and constraint coordinates.

At each target evaluation, the program forms
$\boldsymbol g=\boldsymbol m_g+\boldsymbol B\boldsymbol z$ and
$\mu=m_\gamma+\boldsymbol\lambda^T\boldsymbol z$, then evaluates
the upper threshold $\widetilde\tau_r(\boldsymbol g)$.
In \cref{lst:stan-core}, \texttt{threshold\_upper} denotes this
value, \texttt{conditional\_sd} denotes $s$, and
\texttt{alpha} denotes
$\widetilde\alpha=(\widetilde\tau_r(\boldsymbol g)-\mu)/s$.
The Gaussian inputs already incorporate the observations, so the
model block adds only the log survival weight
\texttt{log\_survival}$=\log\Phi^{\mathrm c}(\widetilde\alpha)$ to the
standard normal log density of $\boldsymbol z$.

External C++ functions supply the threshold and normal-tail operations.
The threshold function follows
\cref{subsec:threshold-computation} and attaches the approximate
envelope derivative from \cref{subsec:threshold-derivatives} to
its returned upper value through a custom automatic differentiation
adapter. Stan can therefore propagate the force through the linear
coefficient map without differentiating the root search or interval
subdivision. The log-survival adapter uses the negative inverse
Mills ratio, and the quantile routine accepts a log probability,
as described in \cref{subsec:tail-computation}.

\begin{listing}[htbp]
\begin{Verbatim}[fontsize=\footnotesize,breaklines]
model {
  target += -0.5 * dot_self(z) + log_survival;
}
generated quantities {
  real eta = -normal_log_quantile(
    log_survival + log(uniform_rng(0, 1)));
  real gamma =
    threshold_upper + conditional_sd * (eta - alpha);
}
\end{Verbatim}
\caption{Marginal sampling and conditional reconstruction in Stan.
The excerpt assumes that the Gaussian transformations, upper threshold,
and log survival weight have been evaluated. The generic routine
\texttt{normal\_log\_quantile} returns $\Phi^{-1}(\exp(\ell))$
from a supplied log probability $\ell$.}
\label{lst:stan-core}
\end{listing}

For each retained $\boldsymbol z$, the generated-quantities block
draws $\gamma$ from its truncated conditional Gaussian law using an
independent uniform variate and the same upper threshold as the model
block. Inverting the coefficient map then gives a joint coefficient
draw. In particular, the monotonicity program restores $a_2=-\gamma$
and $a_4=g_4-\gamma/4$, while the convexity program restores
$a_1=-\gamma$. The other coefficients follow
\cref{tab:coefficient-reconstruction}. The finite-value and quantile
checks in \cref{subsec:tail-computation} apply to this reconstruction.

Extending these programs to a joint restriction requires Gaussian
inputs and reconstruction under the common map in
\cref{cor:combinations}. The external threshold routine must enclose
$\tau_A^\varepsilon$ and supply the envelope force from its active
orders and locations, as described in
\cref{subsec:threshold-computation,subsec:threshold-derivatives}.
With these inputs, the scalar sampling and reconstruction blocks
in \cref{lst:stan-core} apply unchanged.

The returned log values determine the density used to weight NUTS
states, whereas the supplied force determines the proposed trajectory.
To justify this distinction, consider exact arithmetic with a fixed
Euclidean metric, step size, and maximum tree depth after warm-up.
Suppose the log values and force are finite, deterministic, measurable
functions of position and the resulting density is normalizable.
Leapfrog consists of position and momentum shears that preserve volume
and invert when the step size changes sign. Consequently, momentum
reversal makes the proposal reversible even when the force is not the
log-density gradient \citep[Section 5.5]{neal2011hmc}.
With the recursive subtree checks and weighted state selection of
NUTS \citep[Appendices A.3--A.5]{betancourt2017conceptual}, this
preserves the density defined by the returned log values in the ideal
construction without evaluation failures or divergence truncation.
The depth limit makes each transition finite. This argument justifies
using an approximate envelope force but gives no bound on finite-chain
or floating-point error, and the force approximation can still affect
sampling efficiency.

\FloatBarrier
\section{Lidar application details}
\label{app:lidar-computation}

\subsection{Computation}

For the model in \cref{sec:lidar}, computation subtracts the observed
response mean and divides by the response range, applying the same
change of scale to the Gaussian model. The global-threshold tolerance
is $10^{-12}$ in these transformed coefficient units. An expectation
propagation approximation at $129$ derivative locations supplies the
initial mean and dense proposal metric. The continuum threshold
still determines every target evaluation, so these locations affect
only initialization and the metric.

Four Stan chains each use $400$ warm-up iterations and $2{,}000$
retained draws, without thinning. During warm-up, the step size adapts
to target acceptance $0.9$ while the supplied metric is held fixed.
The maximum tree depth is $10$ during warm-up and $12$ during
retained sampling, which otherwise uses the resulting tuning.

\paragraph{MCMC diagnostics.}
We monitor the sampling coordinates and reconstructed coefficients,
the latent response and concentration at the observed ranges, their
adjacent midpoints, and five equally spaced locations, together with
the total response increase, prior quadratic form, and log density.
Across these $1{,}146$ scalar quantities, the maximum rank-normalized
split $\widehat R$ is $1.0034$, and the minimum bulk and tail effective
sample sizes are $2{,}776$ and $2{,}479$, respectively. None of the
$8{,}000$ retained iterations has a divergence or reaches the
tree-depth limit, and the minimum energy Bayesian fraction of missing
information (E-BFMI) across chains is $0.9637$. Three divergences
occurred during warm-up, whose draws were discarded. The retained-draw
diagnostics reveal no warning about mixing in the monitored quantities,
while the predictive checks below assess agreement with the observation
model.

\subsection{Predictive checks}

For each retained monotone posterior curve, define the standardized residuals $e_i(\boldsymbol{a})=(y_i-f_{\boldsymbol{a}}(x(r_i)))/\sigma_i$.
We generate one independent replicate dataset from
\[
 Y_i^{\mathrm{rep}}\mid \boldsymbol{a}\overset{\mathrm{ind}}\sim N(f_{\boldsymbol{a}}(x(r_i)),\sigma_i^2),
 \qquad e_i^{\mathrm{rep}}(\boldsymbol{a})=\frac{Y_i^{\mathrm{rep}}-f_{\boldsymbol{a}}(x(r_i))}{\sigma_i}.
\]
We compute the mean squared residual $n^{-1}\sum_i e_i(\boldsymbol{a})^2$, maximum $\max_i|e_i(\boldsymbol{a})|$, and absolute sample correlation between the paired vectors $(e_1,\ldots,e_{n-1})$ and $(e_2,\ldots,e_n)$, ordered by recorded distance.
Here adjacency refers to consecutive released ranges, whose spacings vary. The two remaining discrepancies are the signed mean residuals over $r_i\leq555$ meters and $r_i>555$ meters, each averaged over the observations in its own half of the range. Applying the same definitions to $e_i^{\mathrm{rep}}$ gives the replicated discrepancies in \cref{tab:lidarcriticism}.

\begin{table}[htbp]
\centering
\begin{tabular}{lrr}
\toprule
Constrained posterior discrepancy & Lower tail & Upper tail \\
\midrule
Mean squared standardized residual & 0.5931 & 0.4069 \\
Maximum absolute standardized residual & 0.3914 & 0.6086 \\
Absolute adjacent residual correlation & 0.5359 & 0.4641 \\
Mean standardized residual at $r\leq555$ m & 0.3526 & 0.6474 \\
Mean standardized residual at $r>555$ m & 0.4235 & 0.5765 \\
\bottomrule
\end{tabular}
\caption[Residual checks for mercury lidar]{
Posterior predictive checks from $8{,}000$ globally monotone posterior curves, with one independent Gaussian replicate dataset per curve.
Lower and upper tails give the fractions of replicated discrepancies at or below and at or above the observed discrepancies for the same posterior curve, with the two range summaries split at the midpoint of $555$ meters.
The summaries condition on the fitted covariance and noise parameters and reuse the fitting data, so they are neither calibrated test probabilities nor out-of-sample predictive scores.}
\label{tab:lidarcriticism}
\end{table}

The upper-tail fractions in \cref{tab:lidarcriticism} range from
$0.4069$ to $0.6474$, with no pronounced disagreement for these
summaries under the specified observation model.

\FloatBarrier

\FloatBarrier
\section{Leaf application details}\label{sec:application-details}

\subsection{Data and computation}

The sequence selected in \cref{sec:application} has sample identifier
$1818$. Of its $16$ raw records, two are manually flagged and removed
by the source study's quality filters. The third excluded record is
the dark measurement, for which insufficient dark adaptation prevents
estimation of dark respiration \citep{rogers2019arctic}.

The marginal-likelihood optimization uses rank $32$ and padding
$L=2$, with the resulting estimates carried into the rank-$24$
model in \cref{eq:application-model}. The constrained fit uses four Stan chains,
each with $1{,}500$ warm-up and $3{,}000$ retained iterations, a dense
metric, target acceptance $0.99$, maximum tree depth $12$, and threshold
tolerance $10^{-8}$. The unconstrained fit uses $20{,}000$ independent
Gaussian posterior draws.

\paragraph{MCMC diagnostics.}
We assess mixing across all $80$ scalar outputs of the constrained fit,
including the coefficient coordinates, reconstructed coefficients,
log density, and quantities used in threshold evaluation and
reconstruction. Across these outputs, the maximum rank-normalized
split $\widehat R$ is $1.0007$, and the minimum bulk and tail effective
sample sizes are $5{,}616$ and $7{,}866$, respectively. None of the
$12{,}000$ retained iterations has a divergence or reaches the
tree-depth limit, and the minimum energy Bayesian fraction of missing
information (E-BFMI) across chains is $0.9642$. These diagnostics
reveal no warning about mixing in the monitored quantities. The
predictive checks below separately assess agreement with the
observation model.

\subsection{Curvature inference}

\begin{figure}[htbp]
\centering
\includegraphics[width=\textwidth]{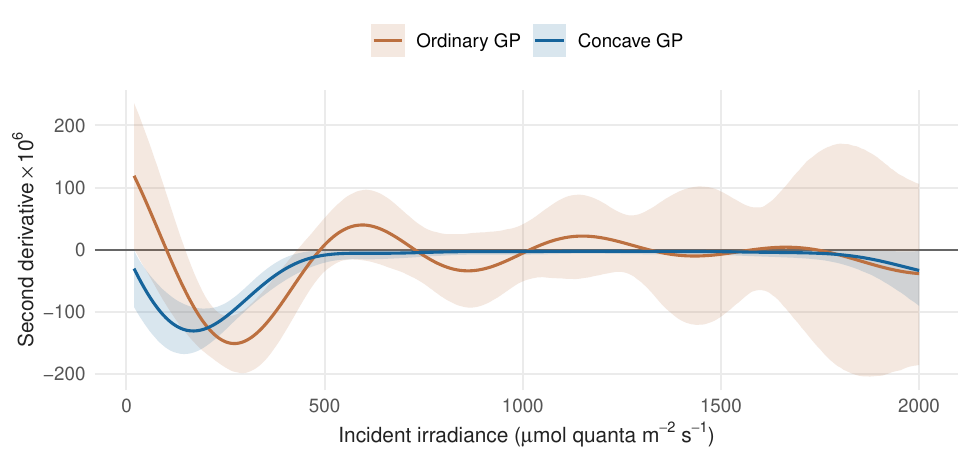}
\caption[Leaf response curvature]{Second derivative of net assimilation
with respect to irradiance for the selected leaf, multiplied by $10^6$,
with unscaled units of assimilation per squared irradiance.
Orange and blue lines show ordinary Gaussian and concavity-conditioned
posterior means and pointwise $95\%$ equal-tail intervals under the same
covariance and noise estimates. The intervals describe latent curvature,
exclude measurement noise, and do not give simultaneous coverage,
while the horizontal reference marks zero.}
\label{fig:curvature}
\end{figure}

To complement the marginal-gain comparison in
\cref{fig:light-response}, \cref{fig:curvature} shows how rapidly
marginal gain changes with irradiance. Concavity restricts the sign of this second
derivative while allowing its magnitude to vary across the range.
The pointwise intervals describe uncertainty about that local rate
of change.

\subsection{Predictive checks}

For each posterior curve from either fit, we generate one independent
replicate dataset at the observed irradiances using the Gaussian
observation model in \cref{eq:application-model}. Define
$r_i(A)=(y_i-A(q_i))/\sigma_\varepsilon$ and let $\pi$ order
the measurements by time. The two discrepancies are the maximum
absolute standardized residual and the absolute sum of adjacent
residual products,
\[
 D_{\max}(\boldsymbol y,A)=\max_i|r_i(A)|,\qquad
 D_{\mathrm{seq}}(\boldsymbol y,A)=
 \left|\sum_{i=1}^{n-1}r_{\pi(i)}(A)r_{\pi(i+1)}(A)\right|.
\]
\begin{samepage}
For either discrepancy $D$, we average the indicator of a replicated
value at least as large as the observed value over all curves from
the corresponding fit. This upper-tail fraction estimates
$P(D(\boldsymbol Y^{\mathrm{rep}},A)\geq
D(\boldsymbol y,A)\mid\boldsymbol y)$, with observed and replicated
discrepancies evaluated using the same posterior curve.
\end{samepage}

The upper-tail fractions for $D_{\max}$ are \AppPpcMaximum\ under
concavity and \OrdPpcMaximum\ without the restriction, while
$D_{\mathrm{seq}}$ gives \AppPpcSequence\ and \OrdPpcSequence,
respectively. Neither comparison shows pronounced disagreement,
although $13$ records provide limited information about dependence
or changes in measurement variance.
Because irradiance decreases with time, the sequence discrepancy
cannot distinguish temporal dependence from systematic response error
across irradiance. Both fractions reuse the fitting data and therefore
are not calibrated test probabilities.

\end{document}